\documentclass[twocolumn,showpacs,prb,amsmath,amssymb,floatfix]{revtex4}
\usepackage{placeins}
\usepackage{epsfig}
\usepackage{amsmath}
\usepackage{amsfonts}
\usepackage{amssymb}
\usepackage{amsbsy}
\usepackage{mathrsfs}
\usepackage{graphicx}% Include figure files
\usepackage{subfigure}
\usepackage{dcolumn}% Align table columns on decimal point
\usepackage{bm}% bold math

\begin{document}

\title{Capacitative coupling of singlet-triplet qubits in different inter-qubit geometries}

\author{Tuukka Hiltunen}
\author{Ari Harju}

\affiliation{ 
  Department of Applied Physics,
  Aalto University School of Science, P.O. Box 14100, 00076 Aalto, Finland
}

\date{\today}

\begin{abstract}
In the singlet-triplet qubit architecture, the two-qubit interactions required in universal quantum computing can be implemented by capacitative
coupling, by exploiting the charge distribution differences of the singlet and triplet states.
The efficiency of this scheme is limited by decoherence, that can
be mitigated by stronger coupling between the qubits.
In this paper, we study the capacitative coupling of singlet-triplet qubits in different geometries of the two-qubit system. The effects of the qubit-qubit distance
and the relative orientation of the qubits on the capacitative coupling strength are discussed using an accurate microscopic model and exact diagonalization of it. We find that the trapezoidal quantum dot formations
allow strong coupling with low charge distribution differences between the singlet and triplet states. The analysis of geometry on the capacitative coupling is also extended to the many-qubit case and the creation of
cluster states. 
\end{abstract}

\maketitle

\section{Introduction}

Two-electron spin eigenstates in semiconductor double quantum dots (DQD) were proposed as qubits \cite{Loss98} by Levy in 2002\cite{levy} and allow a scalable architecture for quantum computation\cite{taylor}.
The universal set of quantum gates for two spin singlet-triplet DQD-qubits has been demonstrated
experimentally\cite{petta2, foletti, shulman}. In
this architecture, the two qubit operations required for universality are implemented using long-distance capacitative coupling by the Coulomb-interaction,
in which the charge asymmetries of the singlet and triplet states are exploited\cite{stepa,taylor,shulman}.
The capacitative coupling of singlet-triplet qubits, resulting in a two-qubit CPHASE-gate, has been demonstrated experimentally\cite{shulman}.

The capacitative two-qubit operation
can be used to create a maximally entangled Bell-state between singlet-triplet qubits. Entanglement is an essential resource in quantum information technology and
is at the heart of all quantum computing\cite{ent}. Coupling more than two qubits together allows
the generation of multipartite entangled states, including the Greenberger-Horne-Zeilinger (GHZ)\cite{greenberger,dur,coff} and
cluster states\cite{briegel,cluster2}. These highly entangled states have applications for example in the proposed one-way quantum computer\cite{one-way2,one-way}, an
alternative to the circuit model of quantum computing\cite{circuit}.

In implementing the qubit operations, the coupling between the quantum dot and the semiconductor environment leads to the common problem in quantum computing, namely decoherence. The most important
electron spin decoherence sources considering singlet-triplet qubits are the coupling to the nuclear spins in host materials such as GaAs \cite{johnson,folk,khaet,merkulov} and the 
the effects of the fluctuating charge environment \cite{hu1,spectro,weperen}. In the experimental realizations of quantum gate operations, the effects of decoherence
can be minimized by decreasing the gate operation times. In the case of the two-qubit capacitative gate, this is can be achieved by enhancing the qubit-qubit coupling.
Stronger coupling also allows the use of smaller charge asymmetries that are less susceptible to the charge noise\cite{ramon,ramon2,requ,weperen}.

In this paper, we model the capacitative coupling of singlet-triplet qubits using accurate exact diagonalization (ED) technique. We study different qubit geometries and 
find that the optimal ones are those in which the quantum dots of the qubits form a trapezoid. These quantum dot formations allow stronger
couplings and hence more efficient quantum gate architectures.
We also find certain 'dead angles' geometries, in which the capacitative coupling disappears completely.
The analysis on the effects of qubit-qubit geometry is also extended to the many-qubit case, where we use an accurate microscopic model to simulate the creation
of cluster states between singlet triplet-qubits.

This paper is organized as follows. In Sec. II, we discuss the computational methods and the simulation model used in the paper. Sec. III is devoted to
the analysis of the two-qubit coupling. We study the effect of the distance of the qubits and their relative orientation on the strength of the capacitative coupling and
the entangling properties of the two-qubit gate. We find the geometries yielding strong qubit-qubit coupling and also analyze the two-qubit gate operation
in different coupling geometries. In Sec. IV, we simulate the creation of cluster states by CPHASE-operations between adjacent qubits.

\section{Model and methods}

\subsection{Continuum model}

A lateral GaAs quantum dot system with $N$ electrons is described with the Hamiltonian
\begin{equation}
\label{eq:Hamiltonian}
H=\sum_{j=1}^N\left[-\frac{\hbar^2}{2m^*}\nabla_j^2+V(\mathbf{r}_j,t)\right]+\sum_{j<k}\frac{e^2}{4\pi\epsilon r_{jk}},
\end{equation}
where $m^*=0.067\,m_e$ and
$\epsilon=12.7\,\epsilon_0$ are the effective electron mass and
permittivity in GaAs, respectively.
The external potential $V(\mathbf{r})$ for quantum dot systems is approximated with a piecewise parabolic potential
that consists of several parabolic wells. A confinement potential of $n$ parabolic wells can be written as
\begin{equation}\label{eq:qdpot}
V(\mathbf{r})=\frac{1}{2}m^*\omega_0^2\min_{1\leq m \leq n}\{|\mathbf{r}-\mathbf{R}_m|^2\}+V_d(t,\mathbf{r}),
\end{equation}
where $\{\mathbf{R}_m\}_{1\leq j\leq n}$ are the locations of the minima of the parabolic dots, and $\omega_0$ is the
confinement strength. A time dependent detuning potential $V_d(t,\mathbf{r})$ is included. 

In our ED computations, the electrostatic detuning between the two minima of a DQD system is modeled as a step function that assumes constant values at each dot. The discontinuity
in the detuning potential is found to have no effects compared to the continuous case of our previous study\cite{leak}. 
The detuning of a singlet-triplet qubit is defined as the potential energy difference between the two parabolic minima of the qubit, i.e.
if the qubit consists of parabolic wells at $\mathbf{R}_1$ and $\mathbf{R}_2$, the detuning is given as $\epsilon(t)=V(\mathbf{R}_1,t)-V(\mathbf{R}_2,t)$.

\subsection{Lattice model}

Although accurate, the continuum model ED is computationally very expensive and thus limited to small particle numbers (no more than two $S-T_0$ qubits can be modeled accurately) and a full
scanning of various QD system geometries is not possible.
A more flexible method for studying systems of several
$S-T_0$ qubits is the extended Hubbard model with the inclusion of a long-range Coulomb interaction.

In this model, a system consisting of $N_q$ singlet-triplet qubits ($N=2N_q$ electrons and QDs) can be described using the Hamiltonian,
\begin{eqnarray}\label{eq:hham}
H&=&\sum_{i\sigma}E_{i\sigma}a^{\dagger}_{i\sigma}a_{i\sigma}\nonumber \\ &-&\sum_{ij\sigma}t_{ij\sigma}a_{i\sigma}^{\dagger}a_{j\sigma} 
+\sum_{ij}U_{ij}n_in_j,
\end{eqnarray}
where $i$ and $j$ are the site indices, and $\sigma$ the spin index. $E_{i\sigma}$ are the on site energies at each QD, $t_{ij\sigma}$ the tunneling element between dots, and $U_{ij}$ the Coulomb-interaction
between sites $i$ and $j$, and $n_i$ the charge at site $i$. In this paper, the tunneling $t_{ij\sigma}=t_{ij}$ is non-zero only between the adjacent dots inside a qubit (i.e. there is no tunneling between the qubits).
The electron-electron interaction is long-range, and is given as
\begin{equation}
U_{ij}=\left[(1-\delta_{ij})\frac{C}{|\mathbf{r}_i-\mathbf{r}_j|-d}+\delta_{ij}U\right],
\end{equation}
where $C=e^2/4\pi\epsilon_r\epsilon_0$ is the Coulomb-strength, $\mathbf{r}_i$ and $\mathbf{r}_j$ are the locations of the dots $i$ and $j$. $U$ is the on-site interaction between two electrons in the same QD and $d>0$ is an extra constant
conveying the fact that in truth the wave functions have finite widths. 
The parameters of the Hubbard model ($U$, $t_{ij}$, and $d$) can be fitted to continuum model data in order to produce more realistic results. 

\subsection{Computational methods}

The continuum Hamiltonian (\ref{eq:Hamiltonian}) is diagonalized using the ED method. In the ED many-body calculations, the one-particle basis is the eigenstates corresponding
to the confinement potential (\ref{eq:qdpot}). The multi-particle basis, in which the Hamiltonian of Eq. (\ref{eq:Hamiltonian}) is diagonalized, is constructed from the single-particle eigenstates as the antisymmetrised Fock states.
The one-particle eigenstates $\{|\psi_p\rangle\}_{p=1}^{N_1}$ (the eigenbasis size being $N_1$) are computed using the multi-center Gaussian basis $\{|\phi_i\}_{i=1}^{N_g}$ (the method is described in detail by Nielsen et al. \cite{requ}).
The Coulomb-interaction matrix elements $V_{i,j,k,l}=\langle\phi_i|\langle\phi_j|V_{int}|\phi_l\rangle|\phi_k\rangle$
can be computed analytically in the Gaussian basis. The elements $V_{i,j}=\langle\phi_i|V(\mathbf{r})|\phi_j\rangle$ can also be computed analytically
for certain confinement potentials $V(\mathbf{r})$, but generally they must be obtained using numerical integration.
The matrix elements $\tilde{V}_{p,q}$ and $\tilde{V}_{p,q,r,s}$ corresponding to the one-particle eigenstates are then computed
from the Gaussian elements by basis changes, as $\tilde{V}_{p,q}=\sum_{i,j}\langle\psi_s|\phi_i\rangle\langle\phi_j|\psi_q\rangle V_{i,j}$ and
$\tilde{V}_{p,q,r,s}=\sum_{i,j,k,l}\langle\psi_p|\phi_i\rangle\langle\psi_q|\phi_j\rangle\langle\phi_k|\psi_r\rangle\langle\phi_l|\psi_s\rangle V_{i,j,k,l}$ (the sums go from $1$ to $N_g$).

In the computation of the one-particle eigenstates, $\{|\psi_p\rangle\}_{p=1}^{N_1}$, an evenly spaced grid of several hundred Gaussian functions (up to $N_g=500$) is used.
The grid dimensions and the Gaussian widths are optimized and the convergence of the states is verified by comparing the energies to ones obtained with a much larger grid.
We perform the basis change corresponding to the elements $\tilde{V}_{p,q,r,s}$ with an Nvidia Tesla C2070 graphics processing unit, which was programmed with CUDA \cite{CUDA}, a parallel programming model for Nvidia GPUs.
The many-body eigenstates are computed with ED using 18 first single-particle states ($N_1=18$). This basis size is found to be sufficient for the convergence of
the results (the relative difference of the many-body ground state energies with 18 and 24 single particle states is less than $0.1\%$ up to very high detuning region).

The continuum Hamiltonian is diagonalized using the Lanczos algorithm for sparse matrices. In the Lanczos method, only the ground state and its energy is obtained accurately. The higher
lying eigenstates can be obtained using a 'ladder operation'. The $k$th state $|\psi_k\rangle$ is obtained as the ground state of the Hamiltonian
\begin{equation}
H_k=H+\delta\sum_{s=1}^{k-1}|\psi_s\rangle\langle\psi_s|,
\end{equation}
where $H$ is the original Hamiltonian of the system and $\delta>0$ is a penalizing constant that moves the lower eigenstates $\{|\psi\rangle_s\}_{s=1}^{k-1}$ above the desired $k$th state. 
The lattice Hamiltonian of Eq. (\ref{eq:hham}) can be diagonalized directly, as its linear dimensions do not exceed $100$ in the computations done in this study. 

The time evolution of a $S-T_0$ qubit system, described by the wave function $|\psi(t)\rangle$ and governed by the Hamiltonian $H(t)$, is computed by propagation,
\begin{equation}
|\Psi(t+\Delta t)\rangle=\exp\left(-i\Delta tH(t)/\hbar\right)|\Psi(t)\rangle.
\end{equation}
Here, $t$ and $\Delta t$ are time and time step length, respectively. $H(t)$ is either the lattice or the continuum Hamiltonian. In the continuum case, the 
exponential is computed using the Lanczos method.

\section{Two capacitatively coupled qubits}
\label{sec:two}
In the capacitative coupling of singlet-triplet qubits, the inter-qubit operations are achieved by exploiting the differences of the charge distributions
of the singlet and triplet states under exchange interaction. With non-zero exchange, achieved by electrically detuning the qubits,
the singlet state localizes more into the dot with lower potential, i.e. the lowest singlet state is a superposition of the symmetric charge state $|S(1,1)\rangle$ and
localized charge state $|S(0,2)\rangle$. The triplet, however, stays in the $(1,1)$ charge configuration due to its spatially antisymmetric wave function.
As the singlet and triplet states have different charge distributions, the Coulomb repulsion between two neighboring qubits depends on their states. This creates an entangling two-qubit CPHASE-gate when
two qubits are detuned simultaneously towards the $|S(0,2)\rangle$ regime.

\begin{figure}[!ht]
\vspace{0.3cm}
\includegraphics[width=0.85\columnwidth]{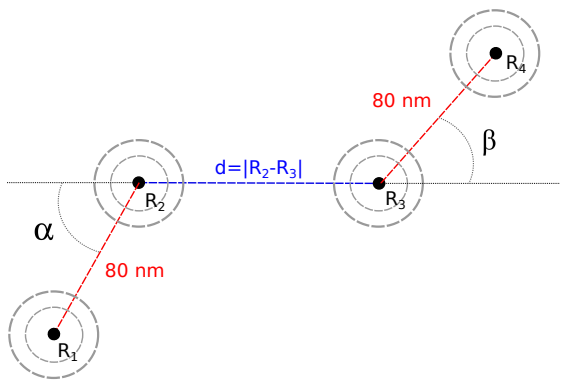}
\caption{(Color online) Locations of the QDs of a two-qubit system. Qubit A consists of dots at $\mathbf{R}_1$ and $\mathbf{R}_2$ and B of those at $\mathbf{R}_3$ and $\mathbf{R}_4$.
The qubit-qubit distance is $d=|\mathbf{R}_2-\mathbf{R}_3|$. The confinement strength is $\hbar\omega_0=4$ meV and the intra-qubit dot distance is $80$ nm. The angles $\alpha$ and $\beta$ determine the locations of dots $1$ and $4$.}
\label{fig:angles}
\end{figure}

The strongest coupling is achieved when the two qubits $A$ and $B$ are initiated in $xy$-plane of the Bloch sphere, and then evolved under exchange, causing them to entangle.
The entanglement can be characterized by an entanglement measure, such as concurrence\cite{ent}. Concurrence $C$ assumes values between $0$ and $1$, and the bigger the value, the stronger the entanglement.

A formula for the evolution of the concurrence by capacitative coupling of two $S-T_0$ qubits can be derived
by writing the
Hamiltonian in the two-qubit basis $\left\{|SS\rangle,|ST_0\rangle,|T_0S\rangle,|T_0T_0\rangle\right\}$, which results in a diagonal matrix with the energies of the aforementioned qubit basis states as its
diagonal entries. In this basis, the two-qubit wave function is written as $|\Psi(t)\rangle=\sum_{X=S,T_0}\sum_{Y=S,T_0}\alpha_{XY}(t)|XY\rangle$, and at $t=0$ all coefficients $\alpha_{XY}=1/2$ as the qubits are
initiated in the $xy$-plane. In the time evolution of the system, each of the four terms obtains a phase factor corresponding to its energy. Defining a $2\times2$-matrix,
$\mathbf{M}(t)$, so that $M_{11}(t)=\alpha_{SS}(t)$, $M_{22}(t)=\alpha_{T_0T_0}(t)$, $M_{12}(t)=\alpha_{ST_0}(t)$, and $M_{21}(t)=\alpha_{T_0S}(t)$, the concurrence is given as
$C(t)=2|\det(\mathbf{M})|$,\cite{ent} yielding
\begin{equation}\label{eq:conc}
C(t)=\frac{1}{2}\sqrt{2-2\cos\left(E_{cc} t/\hbar\right)},
\end{equation}
with the differential cross capacitance energy between the two double-dot systems,
\begin{equation}\label{eq:delta}
E_{cc}=|E_{SS}+E_{T_0T_0}-E_{ST_0}-E_{T_0S}|.
\end{equation}
Here, $E_{SS}$ is the energy of the qubit basis state $|SS\rangle=|S\rangle_A\otimes|S\rangle_B$, and similarly for the other terms. 
$E_{cc}$ determines the speed of the gate operation and the frequency of the entanglement oscillations.
At time $t$ when
$tE_{cc}/\hbar$ is an odd multiple of $\pi$, $C(t)=1$ and the maximal Bell-state entanglement is achieved.

The physics of a system of two capacitatively coupled $S-T_0$ qubits (apart from the decoherence effects \cite{khaet,folk,cnoise}) depend essentially on two things, the intra-qubit tunneling (the tunneling between the two dots of the qubit) 
and the coulomb repulsion between the qubits, i.e. how the two qubits
are located with respect to each other. The tunneling strength controls the anti-crossing energy gap of the singlet charge states $|S(1,1)\rangle$ and $|S(0,2)\rangle$. The locations and distance
of the qubits affect both the energy differences of the charge states and the locations of the anti-crossing, i.e. the detuning required for the singlet
to transition from $(1,1)$ to $(0,2)$. 

The topic of this section is the study of the effect of the geometry of the two-qubit system on the capacitative
coupling strength and the entangling properties of the gate. An illustration of the dot locations in a two-qubit system is shown in Fig. \ref{fig:angles}. 
In this system, the confinement strength is set to $\hbar\omega_0=4$ meV and the intra-qubit dot distance is $|\mathbf{R}_1-\mathbf{R}_2|=|\mathbf{R}_3-\mathbf{R}_4|=80$ nm.
The geometry of this two-qubit system is defined by the angles $\alpha$ and $\beta$ and the inter-qubit distance $d$.

When modeling the qubits with piecewise parabolical potentials, the intra-qubit tunneling is determined by the distance and the confinement strengths of the parabolic wells.
In order for the qubit to function properly, the tunneling barrier between the dots has to be high enough that with zero detuning the singlet and triplet states
are approximately degenerate, i.e. the exchange can be set to a very small value. This sets lower bounds for viable confinement strengths and intra-qubit dot distances.
Changing the intra-qubit tunneling does not have a large effect on the capacitative coupling that is governed by the inter-qubit Coulomb repulsion.

The locations of the qubits have a quite complex effect on the behavior of the two-qubit gate. For example, the Coulomb repulsion can either facilitate or inhibit the
charge transitions to $|S(0,2)\rangle$, depending on the locations of the low-detuned dots in the two qubits. If the furthest away dots are detuned to low potential, the transition
to $(0,2)$ can happen with lower detuning as it will decrease the repulsion between the qubits. Next, we are going to study the behavior of the two-qubit system with several
different inter-qubit geometries. The results shown and discussed in this section are obtained using the continuum model unless stated otherwise. We will discuss the contributions of the qubit-qubit distance ($d$ in Fig. \ref{fig:angles})
and qubit orientations ($\alpha$ and $\beta$ in Fig. \ref{fig:angles}) separately.

\subsection{Linear alignment of qubits}

\begin{figure*}[!ht]
\vspace{0.3cm}
\includegraphics[width=\columnwidth]{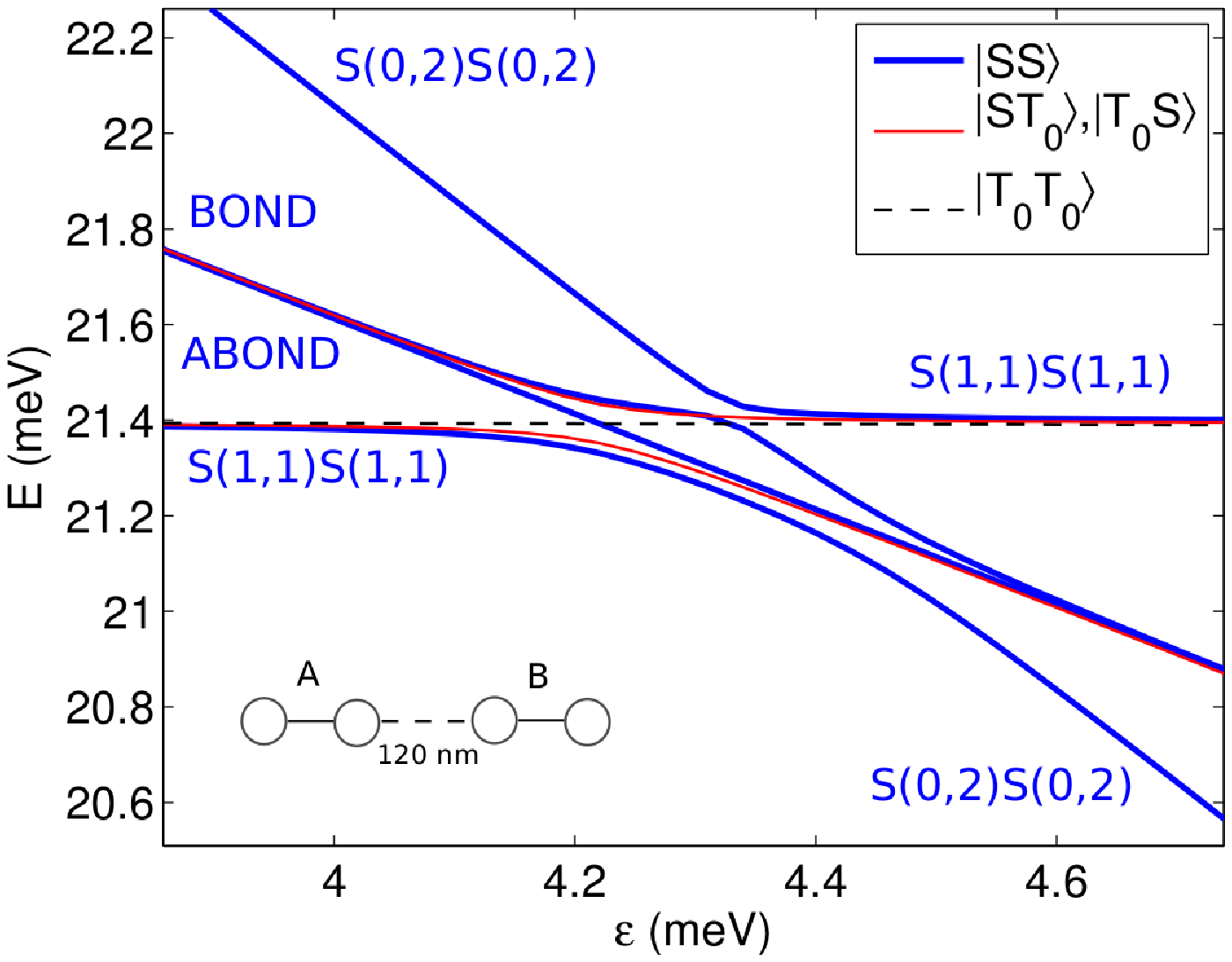}
\includegraphics[width=\columnwidth]{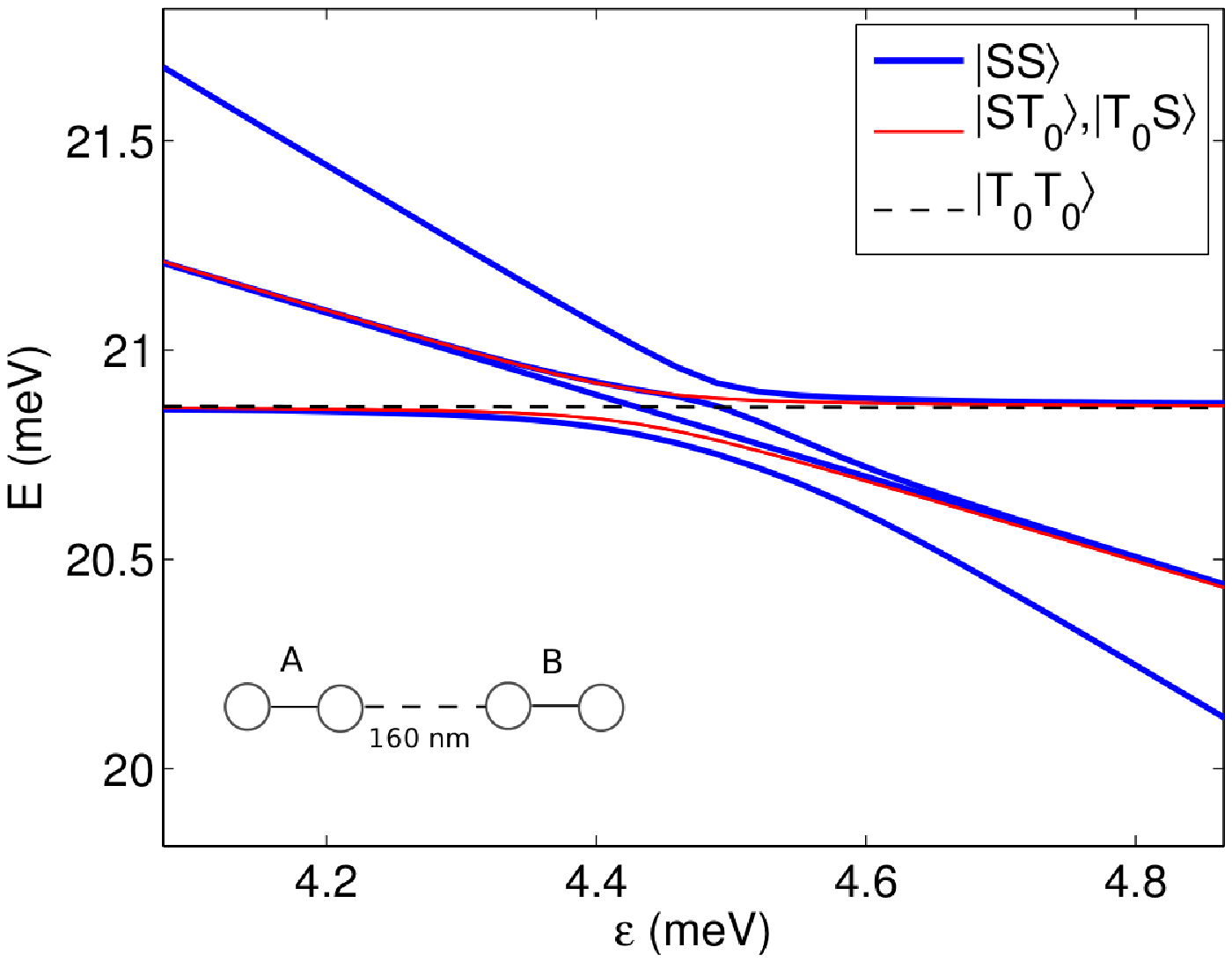}
\includegraphics[width=\columnwidth]{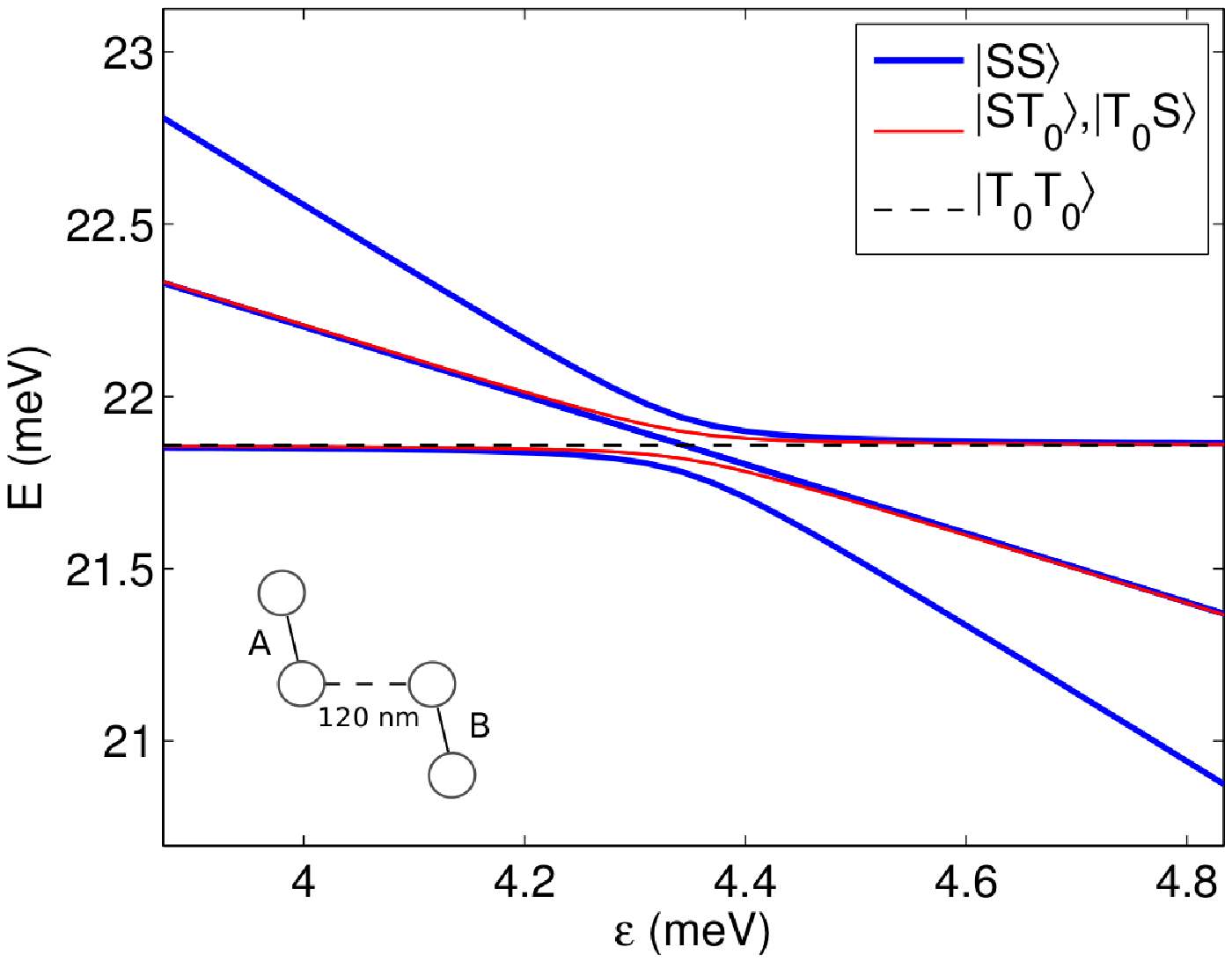}
\includegraphics[width=\columnwidth]{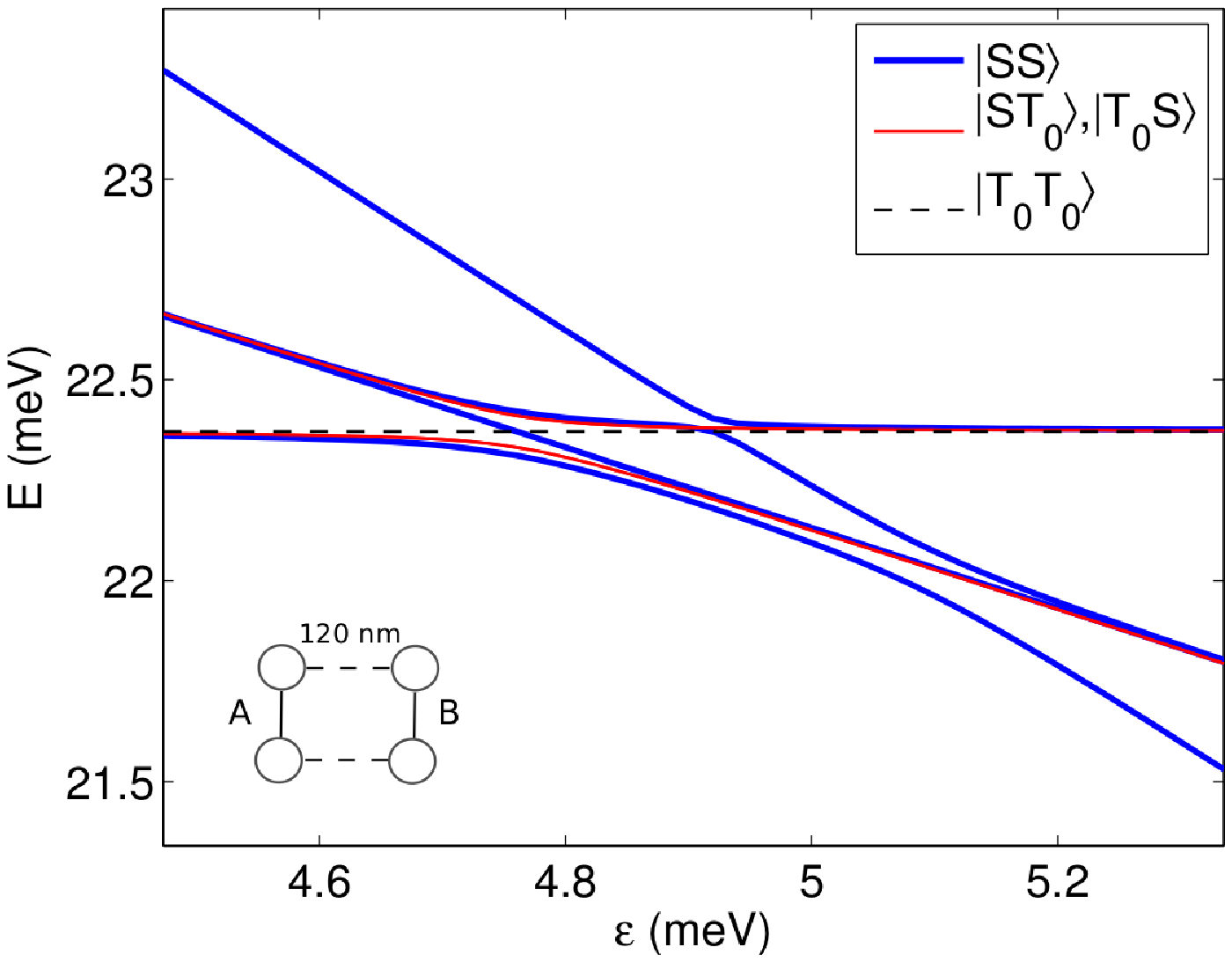}
\caption{(Color online) The energy levels of the two-qubit system as a function of the detuning (see Fig. \ref{fig:angles} for the illustration of the system). Both qubits are in the same detuning $\epsilon_A=\epsilon_B=\epsilon$.
The $|SS\rangle$ states are shown with the thick blue lines, the $|ST_0\rangle$ and $|T_0S\rangle$ states with the red lines, and the $|T_0T_0\rangle$ state with the dashed black line. Illustrations of the corresponding QD formations are shown on top of the plots.
Upper left: ($\alpha=\beta=0$) with the qubit-qubit distance $d=120$ nm. The different singlet-charge states are denoted with labels. Upper right: Linearly aligned system with $d=160$ nm. Lower left: A 'dead angles' case $\alpha=\beta=0.4534\pi$ with $d=120$ nm.
Lower right: Rectangular qubit formation ($\alpha=-\beta=0.5\pi$) with $d=120$ nm.}
\label{fig:ene_vs_det}
\end{figure*}

In the simplest case, both qubits, i.e. all four QDs, are located in a straight line (corresponding to $\alpha=\beta=0$ in Fig. \ref{fig:angles}). 
We study effect of the qubit-qubit distance $d=|\mathbf{R}_2-\mathbf{R}_3|$. The furthest away dots at $\mathbf{R}_1$ and $\mathbf{R}_4$ are detuned
to low potential. The energies of the lowest eigenstates are computed as a function of the detunings $\epsilon_A=V(\mathbf{R}_2)-V(\mathbf{R}_1)=\epsilon_B=V(\mathbf{R}_3)-V(\mathbf{R}_4)=\epsilon$. The energies with $d=100$ nm are shown
in the upper left and the energies with $d=160$ nm in the upper right plot of Fig. \ref{fig:ene_vs_det}. 

The figures show the anti-crossing region of the $|S(1,1)\rangle$ and $|S(0,2)\rangle$ states. 
When the detuning
overcomes the repulsion of two electrons occupying one QD, the ground state singlet shifts from $(1,1)$ to $(0,2)$. In addition to
the $|S(0,2)\rangle_A\otimes|S(0,2)\rangle_B$ and $|S(1,1)\rangle_A\otimes|S(1,1)\rangle_B$ labeled in the figures,
there are also two other $|SS\rangle$-type states (the two blue middle curves besides the $S(0,2)S(0,2)$ and $S(1,1)S(1,1)$), namely 
'the bonding state',
and 'the anti-bonding state',
\begin{equation}\nonumber
\frac{1}{\sqrt{2}}|S(1,1)\rangle_A\otimes|S(0,2)\rangle_B\nonumber
\pm\frac{1}{\sqrt{2}}|S(0,2)\rangle_A\otimes|S(1,1)\rangle_B,
\end{equation}
where $+$ corresponds to the bonding state and $-$ to the anti-bonding state.
The energy eigenstates of a similar two-qubit
system are discussed with more detail in our previous work\cite{leak}, including cases with asymmetric detuning, $\epsilon_1\neq\epsilon_2$.

When the furthest away dots $1$ and $2$ are detuned to low potential, i.e. when $\epsilon_A,\epsilon_B>0$, the Coulomb repulsion caused by the other
qubit facilitates the transition to $(0,2)$; it is preferable for the electrons of qubit A to be as far as possible from qubit B. The shorter the qubit-qubit distance $d$ the larger this effect is.
This is evident from the figures. In the upper left plot of Fig. \ref{fig:ene_vs_det}, the anti-crossing region starts around $\epsilon=3.9$ meV, and in the upper right one around $\epsilon=4.2$ meV. In the $d=100$ nm
case, the eigenenergies are higher due to larger repulsion and the energy differences of the qubit states $\left\{|SS\rangle,|ST_0\rangle,|T_0S\rangle,|T_0T_0\rangle\right\}$ are also affected, and
the width of the anti-crossing region increases as the Coulomb repulsion becomes more prominent. If the closer dots $1$ and $2$ are instead detuned to low potential (i.e. $\epsilon_A<0$ and $\epsilon_B<0$),
the anti-crossing region will shift to higher detuning values, but otherwise the behavior of the two-qubit system stays similar.

The values of $E_{cc}$ (Eq. (\ref{eq:delta})) as a function of the detunings $\epsilon$ with several different qubit-qubit distances are shown in Fig. \ref{fig:delta_ddet}. The value of $E_{cc}$ starts to
increase when the detuning has reached the anti-crossing area, saturating to a constant value when the singlet is fully in $(0,2)$ (it stays constant until the triplet
also starts to undergo the transition to $(0,2)$ at much higher detuning). The shorter the qubit-qubit distance, the larger the maximal value of $E_{cc}$ is, as with shorter
distances the Coulomb repulsion between the qubits has more contribution in the energies of the qubit states. 

\subsection{Rotated qubits}

Next, we discuss the effect of the orientation of the qubits while keeping the qubit-qubit distance constant. The qubits were set $d=|\mathbf{R}_2-\mathbf{R}_3|=120$ nm apart from each other, with the intra-qubit dot distance being $80$ nm. The locations of the furthest away dots
(again detuned to low potential) were varied. An illustration of the system can be seen in Fig. \ref{fig:angles}.
Probing the different values of the angles $\alpha$ and $\beta$ in Fig. \ref{fig:angles} would be very cumbersome using the continuum model. In order to avoid having to compute new sets of the one-particle eigenstates
corresponding to each dot configuration, we study the angle dependence using the Hubbard model of Eq. (\ref{eq:hham}) with its parameters $t_{ij}$, $U$, and $d$ fitted to the continuum model data.

\begin{figure*}[!ht]
\vspace{0.3cm}
\includegraphics[width=\columnwidth]{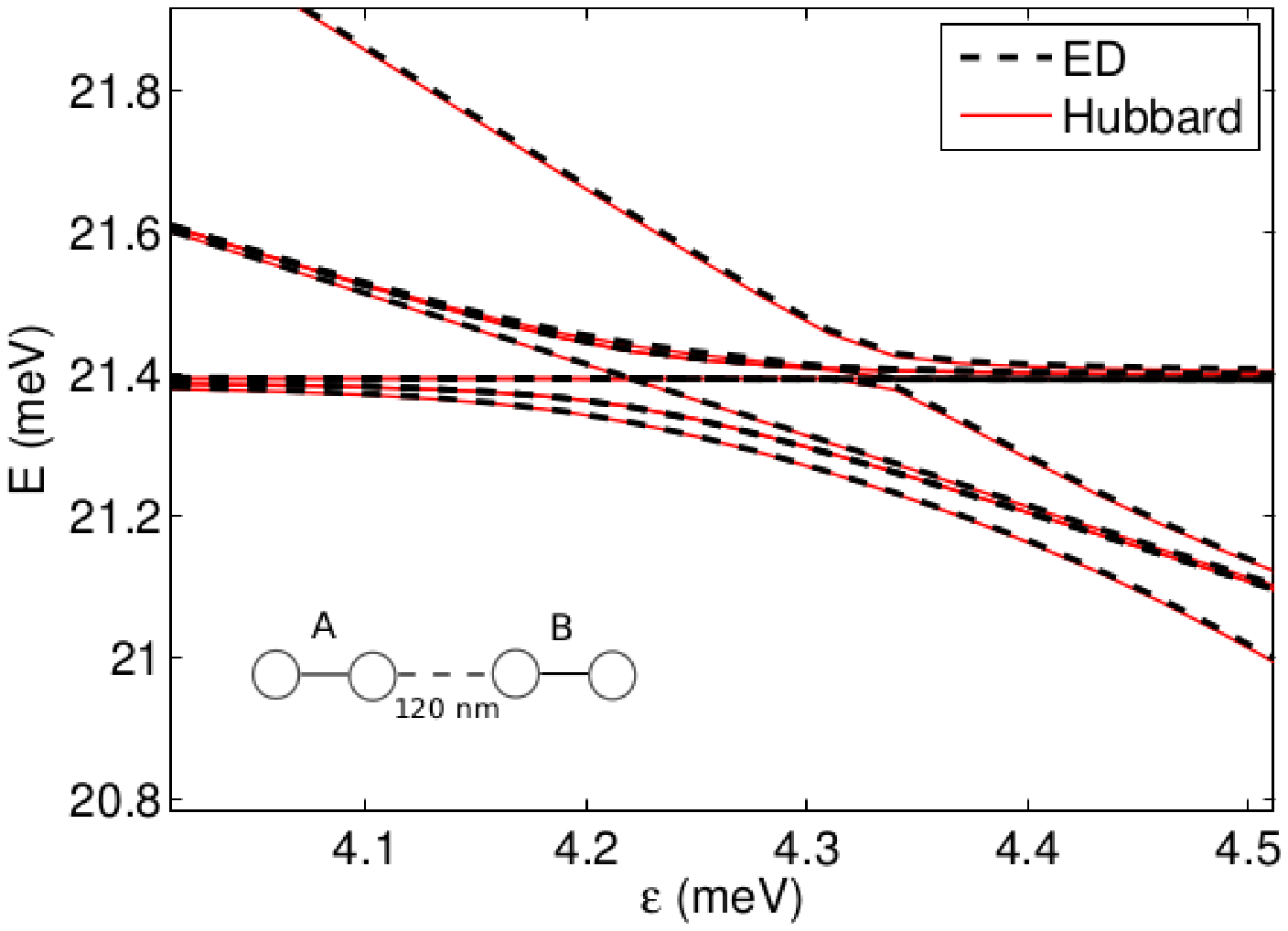}
\includegraphics[width=\columnwidth]{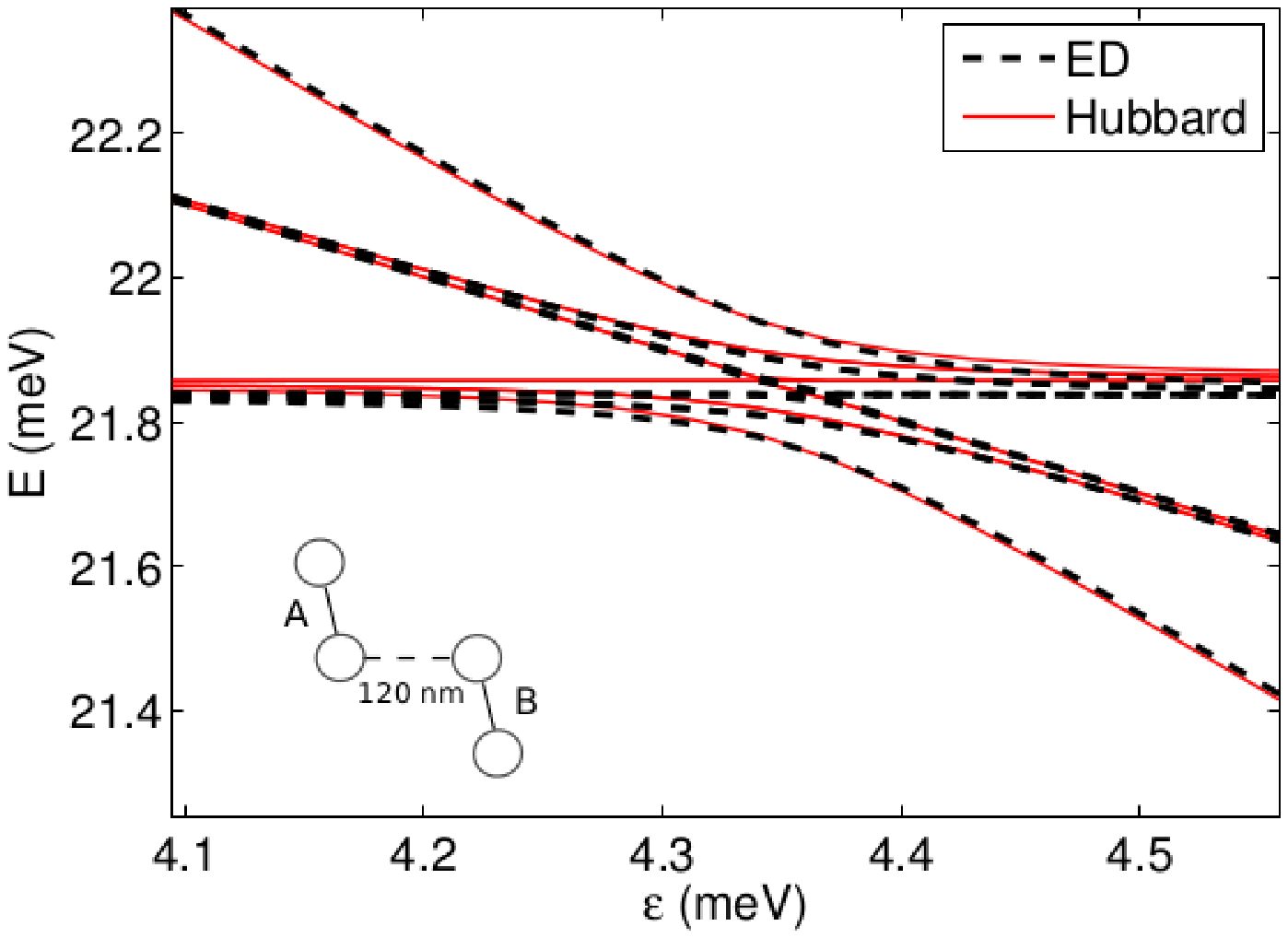}
\caption{(Color online) The lowest energies of a two-qubit system as function of
the detunings $\epsilon_A=\epsilon_B=\epsilon$. The thick black dashed line shows the continuum ED energies, and the
red line the Hubbard energies with the parameters $t_{ij}=27.8$ $\mu$eV, $U=3.472$ meV, and $d=0.43$ nm. These parameters have been obtained by fitting
the Hubbard energies to continuum data corresponding to $\hbar\omega_0=4$ meV confinement, $80$ intra-qubit dot distance, $120$ nm qubit-qubit distance, and linear alignment ($\alpha=\beta=0$ in Fig. \ref{fig:angles}).
The energies of the fitted case are shown in the left panel. The right panel shows the continuum ED and Hubbard energies in the system with $\alpha=\beta=0.4534\pi$ (see Fig. \ref{fig:angles}). The Hubbard parameters are the same as in the left panel, i.e. they are
fitted to the $\alpha=\beta=0$ case.}
\label{fig:ed_vs_hub1}
\end{figure*}

In the case of a reference
system of two $S-T_0$ qubits with the qubit-qubit distance $120$ nm, intra-qubit distance $80$ nm, confinement strength $\hbar\omega_0=4$ meV, and linear alignment, a good fit is obtained
with $t_{ij}=27.8$ $\mu$eV, $U=3.472$ meV, and $d=0.43$ nm.  The parameter fit is demonstrated in the left panel of Fig. \ref{fig:ed_vs_hub1}. The figure shows lowest energies of the two-qubit system (see Fig. \ref{fig:angles} for an illustration of the geometry of the two-qubit system)
as functions of the detunings of the qubits, $\epsilon_A=\epsilon_B=\epsilon$. As seen in the figure, the fitted energies coincide with the continuum model ones. The fit is also tested with asymmetric detunings, $\epsilon_A\neq\epsilon_B$, and is found
equally good in that case. As the capacitative coupling of singlet-triplet is governed by the energy differences of the qubit states, the Hubbard model with the fitted parameters can be used to accurately describe the coupling.

If the geometry of the system (i.e. the dot distances, locations, and the confinements) is changed, generally a new fit for the parameters
has to be computed. However, if the dot distances and the confinement are kept the same and only the qubit orientations (i.e. the angles $\alpha$ and $\beta$ in Fig. \ref{fig:angles}) are changed
the same fit is found to agree well. The right panel of Fig. \ref{fig:ed_vs_hub1} shows the energies of a system with $\alpha=\beta=0.4534\pi$. The Hubbard parameters correspond to the fit
with $\alpha=\beta=0$, i.e. the same parameters used in the left panel of Fig. \ref{fig:ed_vs_hub1}. As seen in the figure, the energies differ now a bit more compared to left panel but
the fit can still be considered good. The Hubbard model can thus be used to probe the values of $\alpha$ and $\beta$ while keeping the qubit-qubit distance constant
 (a task that would be very cumbersome using the continuum model).

\begin{figure}[!ht]
\vspace{0.3cm}
\includegraphics[width=\columnwidth]{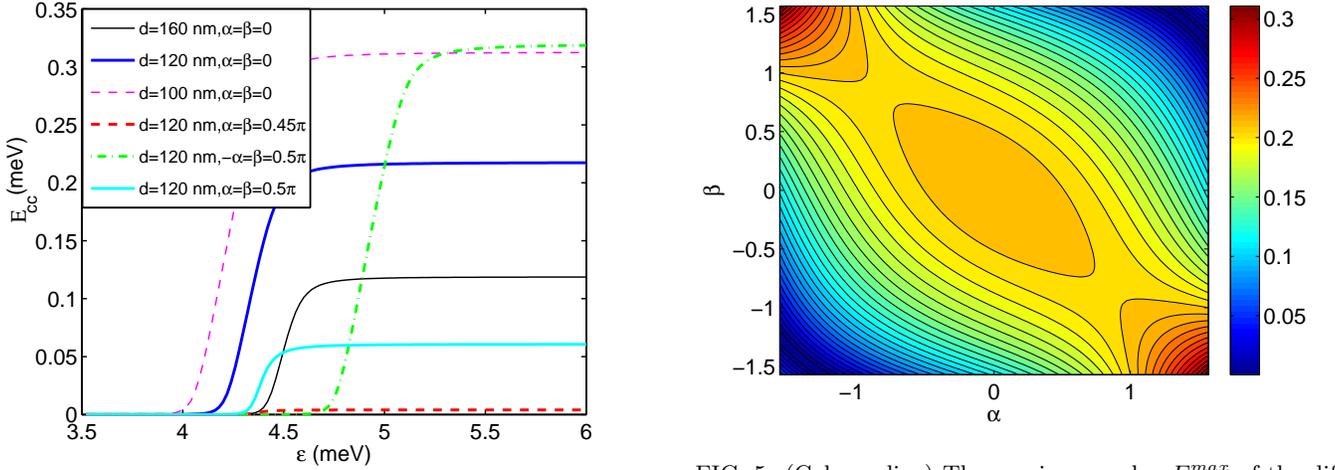}
\caption{(Color online) The values of the cross capacitance $E_{cc}=|E_{SS}+E_{T_0T_0}-E_{ST_0}-E_{T_0S}|$ as functions of the detunings $\epsilon_A=\epsilon_B=\epsilon$
with several different values of the qubit-qubit distance $d=|\mathbf{R}_2-\mathbf{R}_3|$ and angles $\alpha$ and $\beta$ (see Fig. \ref{fig:angles}).}
\label{fig:delta_ddet}
\end{figure}

\begin{figure}[!ht]
\vspace{0.3cm}
\includegraphics[width=\columnwidth]{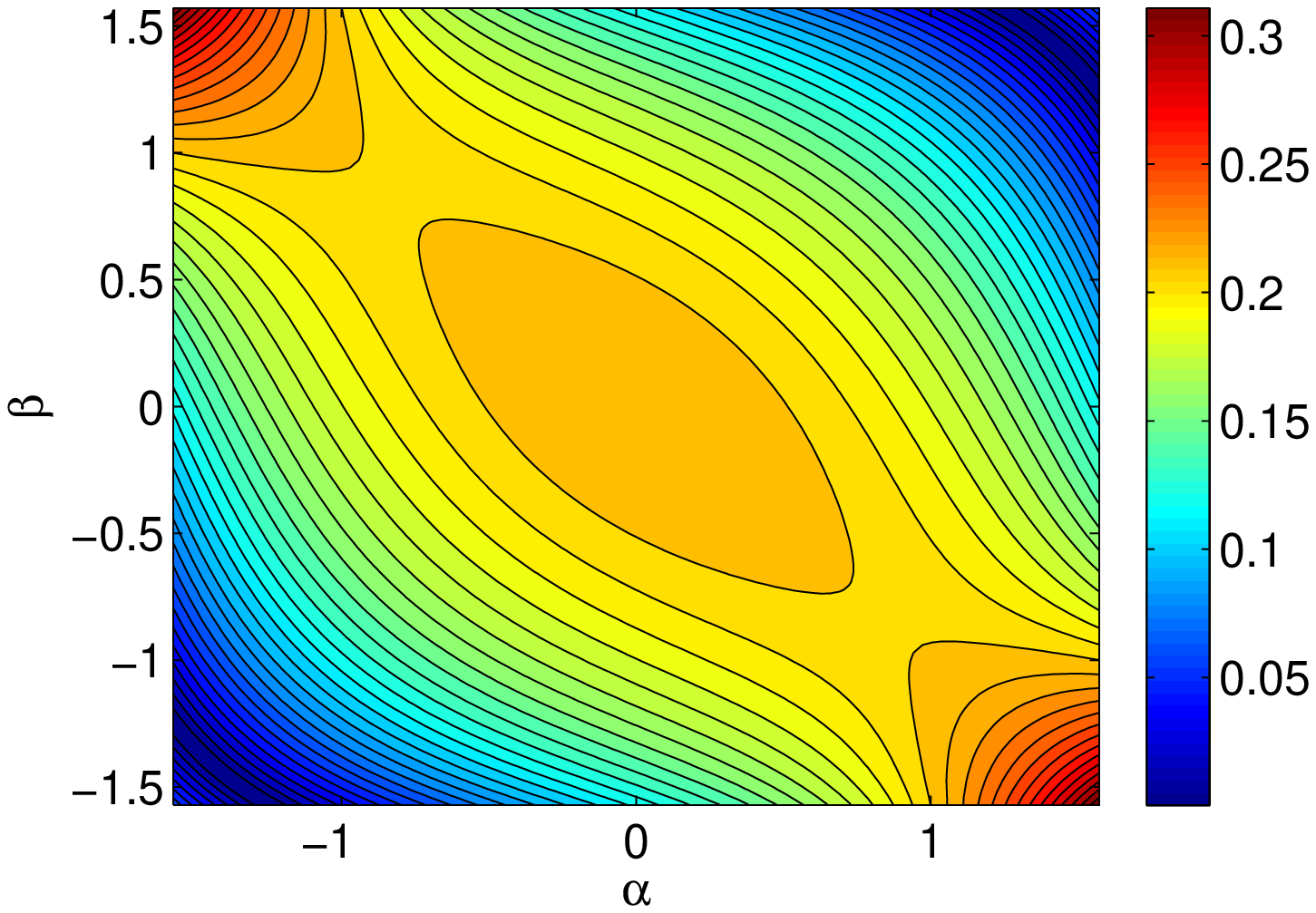}
\caption{(Color online) The maximum value $E_{cc}^{max}$ of the differential cross capacitance energy $E_{cc}$ as a function of the angles
$\alpha$ and $\beta$ (see Fig. \ref{fig:angles} for the description of the system). The $E_{cc}^{max}$ values are shown in meV. The values are computed using
the extended Hubbard model with its parameters fitted to the continuum model data.}
\label{fig:delta_angles}
\end{figure}

The maximum value of $E_{cc}$ as a function
of the angles $\alpha$ and $\beta$ is shown in Fig. \ref{fig:delta_angles}. The figure is obtained by using the Hubbard model of Eq. (\ref{eq:hham}) with the parameters
fitted to the continuum model data for the $\alpha=\beta=0$ case ($t_{ij}=27.8$ $\mu$eV, $U=3.472$ meV, and $d=0.43$ nm).
The Hubbard data with these fitted parameters and continuum model data were also compared with several other angles $\alpha$ and $\beta$, and
the fit was found to be good with arbitrary angles (see Fig. \ref{fig:ed_vs_hub1}).
In Fig. \ref{fig:delta_angles}, the detuning is symmetrical, $\epsilon_A=\epsilon_B$, but the orientations resulting in high coupling applies for the general case as well, i.e. the angle dependence of $E_{cc}$ is similar
also with asymmetric detunings, such as $\epsilon_A=-\epsilon_B$.
Fig. \ref{fig:delta_ddet} shows the $E_{cc}$ values as functions of the detunings $\epsilon$ with several different angles.

As seen in Fig. \ref{fig:delta_angles}, $E_{cc}^{max}$ obtains its largest values along the line $\alpha=-\beta$ that corresponds to the geometries in which the QDs of the system form a trapezoid.
The $E_{cc}^{max}$ values start to increase rapidly when the angles approach the rectangular formation at $\alpha=\pi/2$ and $\beta=-\pi/2$ and with $E_{cc}^{max}=0.3186$ meV. The qubit state energies in this case are shown in the lower right plot of Fig. \ref{fig:ene_vs_det}. 
If the angles are further increased in this direction, the values keep rising but in this case the qubit-qubit distance becomes smaller than $120$ nm.

Along the line $\alpha=\beta$, the values of $E_{cc}^{max}$ decrease as the angles are increased.
Near $\alpha=\beta=\pi/2$, the maximum value has decreased to zero. For example, with $\alpha=\beta=0.4534\pi$ the maximum value is $E_{cc}^{max}=1.7648\times10^{-10}$ meV
compared to the $\alpha=\beta=0$ value $E_{cc}^{max}=0.2159$ meV. With these 'dead angles', the energy difference $E_{SS}+E_{T_0T_0}-E_{ST_0}-E_{T_0S}$ vanishes and $E_{cc}$ is
uniformly zero. The qubit state energies for the dead angle case $\alpha=\beta=0.4354\pi$ are shown in the lower left plot of Fig. \ref{fig:ene_vs_det}. Comparing this to the other plots of the figure,
the anti-crossing area width is smaller, and the intermediate bonding and anti-bonding $|SS\rangle$-states are completely degenerate. 
As the angles are then further increased, the values of $E_{cc}$ start to rise again
reaching $E_{cc}^{max}=0.06$ meV at $\alpha=\beta=\pi/2$. Fig. \ref{fig:delta_ddet} shows the $E_{cc}$-values in the rectangular case $-\alpha=\beta=0.5\pi$, with $\alpha=\beta=0.45\pi$
(close to the dead angle case), and with $\alpha=\beta=\pi/2$.

It should be noted that the behavior of the $E_{cc}^{max}$ as a function of the angles $\alpha$ and $\beta$ or the qubit distance $d$ can be explained electrostatically. $E_{cc}$ obtains its maximum
values when the singlet is fully in the $(0,2)$ configuration. Computing the value of $E_{cc}^{max}$ according to Eq. (\ref{eq:delta}) so that the singlet consists of two unit charges located
at a single dot, and the triplet of one charge per dot, yields the same angle dependence that is seen in Fig. \ref{fig:delta_angles}.

The strength of the qubit-qubit coupling, the energy difference $E_{cc}$, is observable by looking at the energies of the two qubit systems.
The energy difference between the intermediate bonding and anti-bonding states (the width of the 'middle bulges' in Fig. \ref{fig:ene_vs_det})
is determines the strength of the capacitative coupling. 
In the cases where the bonding and anti-bonding states are close to degenerate (small 'bulge' as in Figs. \ref{fig:ene_vs_det}b and \ref{fig:ene_vs_det}c),
$E_{cc}$ assumes very small values, and the qubits are only weakly coupled.

\subsection{Gate operation}

In implementing the capacitative CPHASE-gate, the strongly coupled geometries are preferable.
Long gate operation times mean
more errors due to decoherence from for example the semi-conductor nuclear spin bath \cite{khaet,folk,Jani2} and charge-noise \cite{ramon,spectro,shulman}. In order to achieve fast enough gate operation in the weakly coupled
cases, one needs larger charge distribution differences between the singlet and triplet states. Furthermore, the charge-noise induced decoherence, an important error source in $S-T_0$ qubits, is increased considerably when the qubits are
operated close to the $(0,2)$-regime\cite{taylor,spectro,ramon}. It should be noted that there has also been theoretical studies on additional effects beside the charge asymmetry, so called sweet spots
in capacitative coupling of $S-T_0$, that can minimize the charge noise coupling (e.g. Refs. \onlinecite{Nielsen,ramon2}).

\begin{figure}[!ht]
\vspace{0.3cm}
\includegraphics[width=\columnwidth]{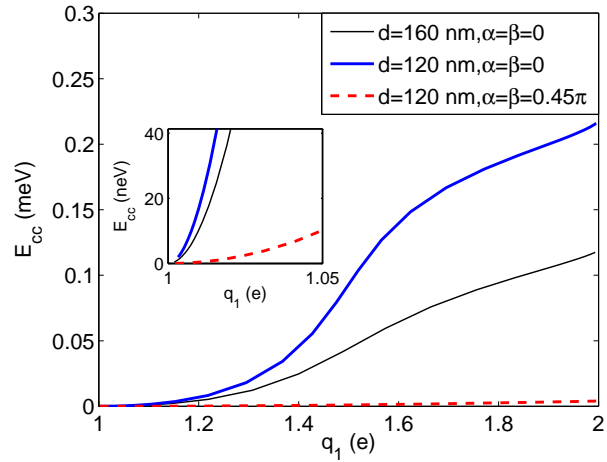}
\caption{(Color online) The cross capacitance $E_{cc}$-values as a function of the charge in the low detuned dot $1$ of qubit A, $q_1$ (the charge in the low detuned dot $4$ of qubit B is exactly the same due to symmetry).
The inset shows at the small $E_{cc}$ region relevant to the CPHASE gate operation time of tens of nanoseconds.}
\label{fig:charge}
\end{figure}

Fig. \ref{fig:charge} shows the fact that with weak coupling one needs large charge asymmetries in order to achieve fast operation. The figure \ref{fig:charge} shows the $E_{cc}$ values as a function of the charge in the low detuned dot of qubit $A$ (the charge in dot $4$, the low detuned QD of qubit B, is exactly the same due to symmetry).
The values are shown in three different geometries: linear with $d=160$, linear with $d=120$ nm with, and $d=120$ nm with $\alpha=\beta=0.45\pi$.
It is evident from Fig. \ref{fig:charge} that in the weakly coupled $\alpha=\beta=0.45\pi$ case, the $E_{cc}$-values
stay small even with very large charge distribution asymmetries. 
In order to achieve the gate operation time of for example $50$ ns (meaning that it takes $50$ ns to achieve the maximal Bell-state entanglement),
a cross capacitance energy of $E_{cc}=41.36$ neV is required (see the inset in Fig. \ref{fig:charge}). In the $\alpha=\beta=0$ case with $d=120$ nm, this corresponds to the charge asymmetry $q_1=1.0161$, and in the $d=160$ case to $q_1=1.0257$. In the weakly coupled $d=120$, $\alpha=\beta=0.45\pi$ case, a much larger asymmetry of
$q_1=1.1203$ e is needed for the same operation.

\begin{figure}[!ht]
\vspace{0.3cm}
\includegraphics[width=\columnwidth]{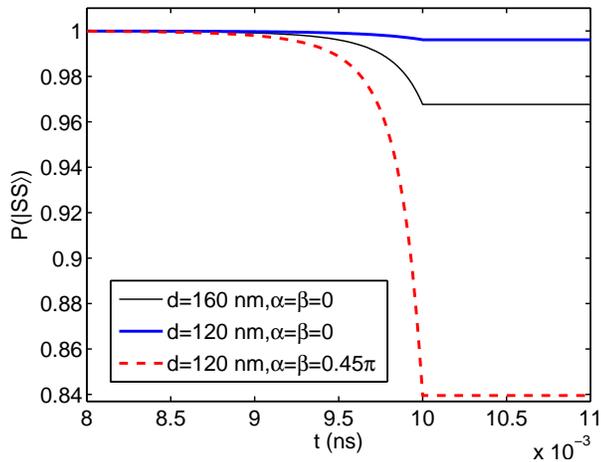}
\caption{(Color online) Occupation of the lowest $|SS\rangle$ states as a function of time. The two-qubit system
is initiated in the $|S(1,1)\rangle_A\otimes|S(1,1)\rangle_B$ charge state. At $t=0$ the detuning sweep is started. The detuning is increased to its maximum value ($\epsilon=3.92$ meV in the $\alpha=\beta=0$ case and $\epsilon=4.25$ meV in 
the $\alpha=\beta=0.45\pi$ case) that corresponds
to a CPHASE-operation with the duration of $50$ ns. The detuning sweep's duration is $\tau=0.01$ ns. At $t=0.01$ ns, when the detunings have reached their final values,
the system is let to evolve for $0.001$ ns.}
\label{fig:leak_angle}
\end{figure}

In addition to the charge-noise induced decoherence, large charge asymmetry can also cause
problems in the form of the singlet charge state leakage. If the singlet is detuned close to the $(0,2)$-regime non-adiabatically, the leakage between $(1,1)$ and $(0,2)$
could hinder the gate operation\cite{leak}. We simulate a non-adiabatic detuning sweep to a charge state corresponding to the aforementioned gate operation
of $50$ ns using the Hubbard model (the parameters are fitted to the continuum model data). The two-qubit system is initiated in $|S(1,1)\rangle_A\otimes|S(1,1)\rangle_B$. The detunings are then increased linearly to their maximum values ($\epsilon=3.92$ meV in the $\alpha=\beta=0$ case, $\epsilon=4.22$ meV in the $d=160$ nm case, and $\epsilon=4.25$ meV in 
the $\alpha=\beta=0.45\pi$ case) during a time of $\tau=0.01$ ns. The occupations of the lowest $|SS\rangle$ and $|ST_0\rangle$ states are plotted as a functions of time in Fig. \ref{fig:leak_angle}.

In the $d=120$ nm and $\alpha=\beta=0$ case, the singlet charge distribution corresponding to the $50$ ns gate operation is very close to $(1,1)$ ($q_1=1.0161$), and
thus the leakage from the lowest singlet state to the higher one is negligible. In Fig. \ref{fig:leak_angle}, the occupations of the lowest $|SS\rangle$ and $|ST_0\rangle$
states are $P(|SS\rangle)=0.996$ at the end of the detuning sweep. With $d=160$ nm, the final occupation is $P(|SS\rangle)=0.968$. In the $\alpha=\beta=0.45\pi$ case, the final probability
is $P(|SS\rangle)=0.840$. The weaker coupling arising from the geometry of the two-qubit system and the consequent large charge asymmetries
required for the gate operation result in probability leaking out of the qubit basis to higher singlet states if the detuning sweep is too fast. In this case, the gate cannot achieve maximal Bell-state
entanglement\cite{leak} (the maximum concurrence is given by the final occupation of the lowest $|SS\rangle$ state). On the other hand, slow detuning pulses needed for an adiabatic passage to $(0,2)$ could
cause problems in controlling the qubits and their interaction.
The non-adiabatic charge state leakage is discussed with more detail in our previous study\cite{leak}.

Our analysis shows that the values of $E_{cc}^{max}$ increase with decreasing qubit-qubit distance, and the largest values are obtained
with trapezoidal dot formations, i.e. with the angles in Fig. \ref{fig:angles} being $\alpha=-\beta$. The strongest coupling (with a given qubit-qubit distance) corresponds
to a rectangular formation of the qubits ($\alpha=\pi/2$ and $\beta=-\pi/2$). In implementing the coupling scheme, systems of large $E_{cc}^{max}$ values are preferable. Weak coupling arising from the geometry (long qubit-qubit distance or the 'dead angles') can
cause problems for example in the form of charge state leakage. It should be noted that the angle dependence of $E_{cc}$ was found to be similar also with asymmetric detunings
and with the closer dots $2$ and $3$ detuned to low potential (i.e. $\epsilon_A,\epsilon_B<0$).

\section{Several qubits and cluster states}

Bell states are created by applying a CPHASE-gate between two capacitatively coupled $S-T_0$-qubits. By applying such gates between all adjacent pairs of an $n$-qubit array, a state belonging
in the higgly entangled class of cluster states is created. A cluster state corresponding to a certain array of qubits can be parametrized by a mathematical graph $G=(V,E)$, with the vertices $V$ being the qubits and the edges
$E$ corresponding to their couplings\cite{briegel}. Cluster states have applications for example in the proposed one-way quantum computing scheme where the system of qubits is first prepared into a cluster state
quantum computing algorithms are then implemented by measuring the qubits in certain order and basis\cite{one-way}.

A general cluster state $|\phi_{N_q}\rangle$ of $N_q$ $S-T_0$-qubits can be written as
\begin{equation}
|\phi_{N_q}\rangle=\frac{1}{2^{N_q/2}}\bigotimes_{k=1}^{N_q}(|S\rangle_k\sigma_z^{k+1}+|T_0\rangle_k),
\end{equation}
where the subscript $k$ corresponds to the qubit number $k$, and $\sigma_z^{k+1}$ is the $z$-Pauli matrix, (here, $\sigma_z^{N_q+1}=1$). In the two-qubit case,
$|\phi_2\rangle=\frac{1}{4}(|SS\rangle-|ST_0\rangle+|T_0S\rangle+|T_0T_0\rangle)$, a state which is equivalent to the maximally entangled Bell states $|\Psi_{\pm}\rangle=\frac{1}{\sqrt{2}}(|SS\rangle\pm|T_0T_0\rangle)$
up to local unitary transformations. Similarly, $|\phi_3\rangle$ is equivalent to the maximally entangled three-qubit Greenberger-Horne-Zeilinger (GHZ) states\cite{greenberger} $|GHZ_{\pm}\rangle=\frac{1}{\sqrt{2}}(|SSS\rangle\pm|T_0T_0T_0\rangle)$ (however,
with $N_q\geq4$, the corresponding cluster states are not equal to the GHZ-states\cite{briegel}).

In generating cluster states, the CPHASE-operations
between qubits can in principle be applied in an arbitrary order, simultaneously or in sequences\cite{briegel,one-way,cluster2}.
In Ref. \onlinecite{maxtri}, we proposed a three-qubit gate for the creation of the three-qubit cluster state (or the GHZ-state) in which the gate operation consists of just one step, setting all three qubits
to the desired detuning values. In this scheme, the qubits were placed symmetrically in a triangular formation. In this section, we simulate the generation of GHZ-states in several other three-qubit geometries.
The following subsections concentrate in detail to two different qubit geometries that we use to demonstrate the both the sequential on and one-step preparation of cluster states.

We simulate three qubits A (consisting of dots $1$ and $2$), B (dots $3$ and $4$), and C (dots $5$ and $6$), with the intra-qubit dot distances being $80$ nm and the qubit-qubit distances $120$ nm. The confinement strength is $\hbar\omega_0=4$ meV. 
The detunings are defined as $\epsilon_A=V(\mathbf{R}_2)-V(\mathbf{R}_1)$, $\epsilon_B=V(\mathbf{R}_4)-V(\mathbf{R}_3)$ and $\epsilon_C=V(\mathbf{R}_5)-V(\mathbf{R}_6)$.
The simulations are done using the extended Hubbard model of Eq. (\ref{eq:hham}). The parameters of the model ($t_{ij}=27.8$ $\mu$eV, $U=3.472$ meV, and $d=0.43$ nm)
are fitted to the two qubit continuum model data with the same dot distances. We characterize the entanglement in three-qubit states using the pairwise concurrences and the three-tangle\cite{coff} that
are computed at each time step as the system is evolved (the six-electron wave function is projected onto the three-qubit computational basis, and the concurrences
are computed as in \cite{coff}). The time-evolution of the six-electron wave function, $|\Psi(t)\rangle$, is computed by propagation with the Hubbard Hamiltonian of Eq. (\ref{eq:hham}),
$|\Psi(t+\Delta t)\rangle=\exp\left(-i\Delta tH(t)/\hbar\right)|\Psi(t)\rangle$.

\subsection{Linear alignment and sequential preparation}

\begin{figure}[!ht]
\vspace{0.3cm}
\includegraphics[width=0.8\columnwidth]{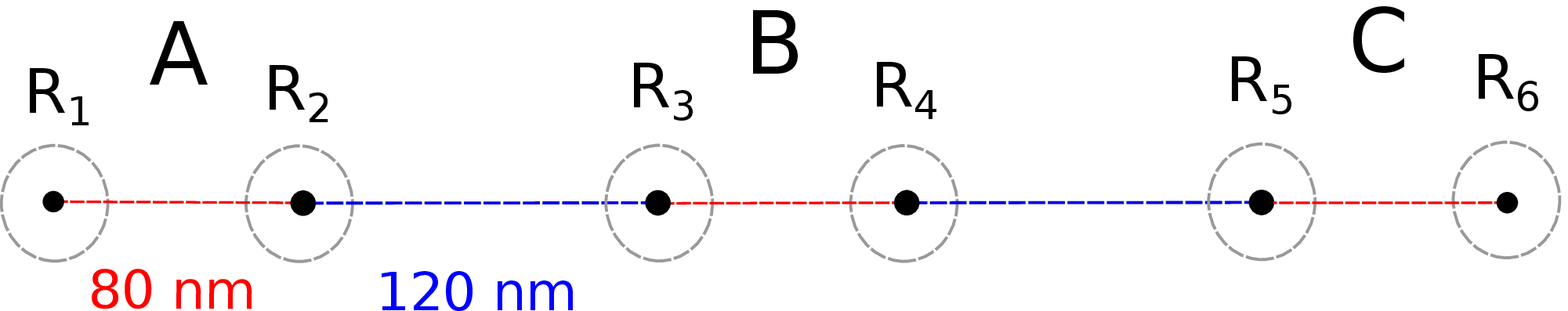}
\includegraphics[width=\columnwidth]{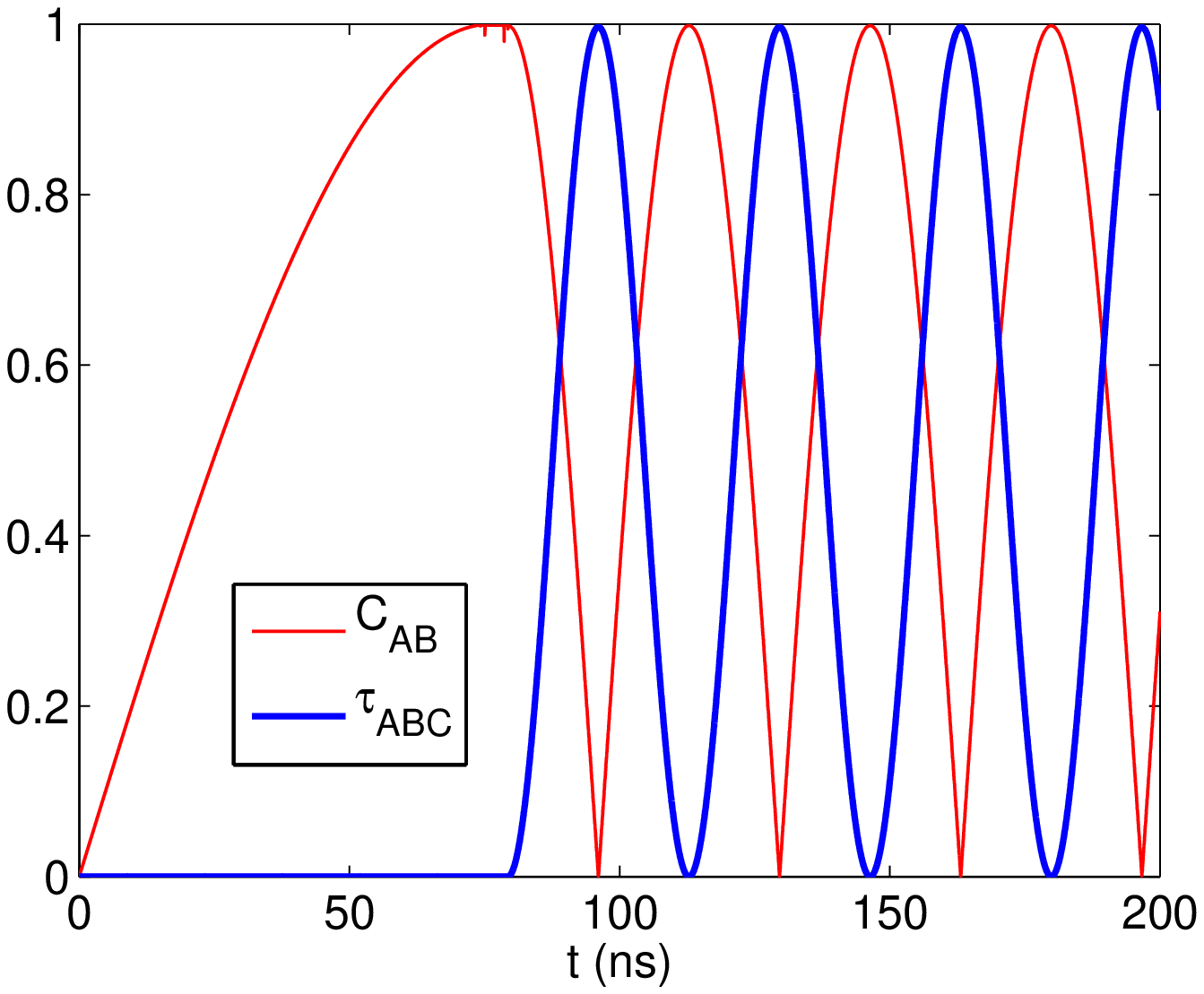}
\caption{(Color online) Evolution of the concurrences in a three-qubit system in the linear alignment geometry shown above the plot. The thick line shows the three-tangle $\tau_{ABC}$ and the thin line the pairwise concurrence of qubits A and B, $C_{AB}$.
The qubits A, B and C are initiated in the state $\frac{1}{\sqrt{2}}(|S\rangle+|T_0\rangle)$ with the detunings $\epsilon_A=-\epsilon_B=4.0$ meV and $\epsilon_C=0$. When $C_{AB}$ has reached the maximum value $1$ at $t=76$ ns,
the detunings $\epsilon_A$, and $\epsilon_B$ are decreased to zero, after which $\epsilon_B$ and $\epsilon_C$ are increased to the values $4.1$ meV and $-4.1$ meV, respectively (the detuning sweeps are adiabatic
with respect to the charge state transition, lasting $2.5$ ns). The system is then let to evolve for $200$ ns.}
\label{fig:cluster}
\end{figure}

In the linear alignment case (see the illustration in Fig. \ref{fig:cluster}), the Coulomb repulsion between the qubits affects the middle qubit B differently
compared to A and C. This asymmetry between the qubits makes it difficult to tune the detunings of the qubits so that the Coulomb repulsion
between them is symmetric. With significantly asymmetric repulsion, entangling the three qubits in a single step is not viable, as
the entanglement oscillates very irregularly in this case (see the asymmetry discussions in Ref. \onlinecite{maxtri}).
Instead, we simulate a more general scheme in principle applicable to any array of $S-T_0$ qubits in which the interactions between qubits are turned on sequentially.

The evolution of the concurrences during the three qubit-operation are shown in Fig. \ref{fig:cluster}. All three qubits A, B, and C are first initiated in the state $\frac{1}{\sqrt{2}}(|S\rangle+|T_0\rangle)$. Qubits A and B are then evolved under exchange with the detunings $\epsilon_A=-\epsilon_B=4.0$ meV.
The pairwise concurrence $C_{AB}$ of A and B is computed, and when it reaches the maximal value $C_{AB}=1$, i.e. when A and B are maximally entangled, the detunings $\epsilon_A$ and $\epsilon_B$ are
decreased to zero adiabatically. Qubits B and C are then detuned to the values $\epsilon_B=4.1$ meV and $\epsilon_C=-4.1$ meV adiabatically (i.e. in this pairwise operation, the furthest away dots
are again detuned to low potential). When the detunings have reached their maximal values,
the system is let to evolve for $200$ ns.

As the qubits A and B are evolved under exchange, they start to entangle, as seen in the increasing values of $C_{AB}$ in the figure. Switching off the detunings of A and B when their subsystem
has reached the maximal Bell-state entanglement, and subsequently evolving B and C under exchange results in true three-qubit entanglement. The system starts to oscillate
between a GHZ-state ($\tau_{ABC}=1$, for example at $t=96$ ns) and a Bell-state between qubits A and B ($C_{AB}=1$). The pairwise concurrences $C_{BC}$ and $C_{AC}$ stay approximately zero in this case;
the entanglement in the system is either of the $GHZ$-type between all three qubits or between just A and B.

The frequency of the oscillations is given by the value of $E_{cc}$ similarly to Eq. (\ref{eq:conc}) (this also
applies to the frequency of $\tau_{ABC}$-oscillations although their functional form is different).
in the two subsystems. In Fig. \ref{fig:cluster}, B and C are detuned to higher values, resulting in larger value of $E_{cc}$ and faster oscillation compared to the beginning of the simulation.

If the detunings of A and B are switched off before (or after) the maximal Bell-state peak, and B and C then detuned to exchange, also $C_{BC}$ achieves non-zero values. In this case, $\tau_{ABC}$ never reaches $1$, and the entanglement oscillates between $C_{AB}$ and $C_{BC}$. For example, if
the switch-off is done at $C_{AB}=0.7$, both $C_{AB}$ and $C_{BC}$ oscillate between $0$ and $0.7$, while $\tau_{ABC}$ assumes values between $0$ and $0.49$.

The linear system was also studied with other qubit-qubit distances including asymmetric cases
where the distance of A and B is different from the distance of B and C. The results were qualitatively similar to the ones already discussed. The geometry of the three-qubit system defines the the $E_{cc}$-values
for the pairwise qubit couplings, which in turn determine the frequencies of the entanglement oscillations. Generally, the results discussed in Sec. III are directly applicable to
the sequential entanglement scheme of arbitrary number of qubits, as in this case the entanglement is generated using only two-qubit CHPASE-operations. 

\subsection{Parallel alignment and one-step preparation}

\begin{figure}[!t]
\vspace{0.3cm}
\includegraphics[width=0.6\columnwidth]{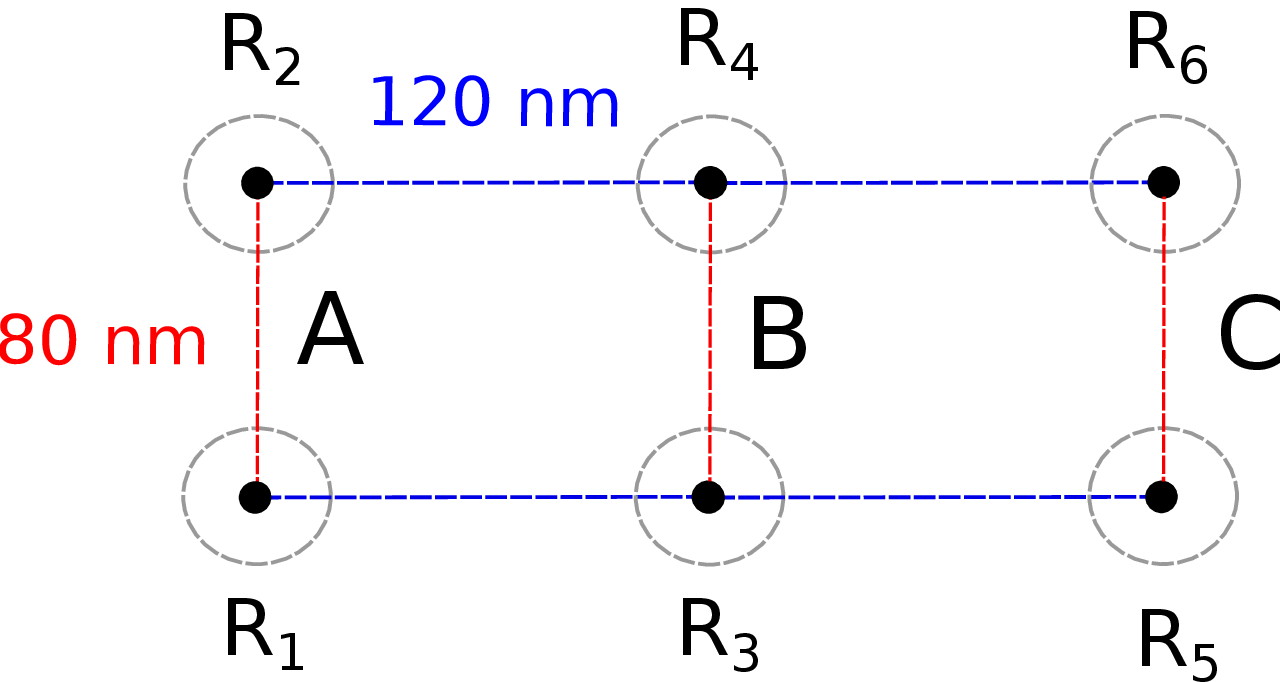}
\subfigure{\includegraphics[width=.48\columnwidth]{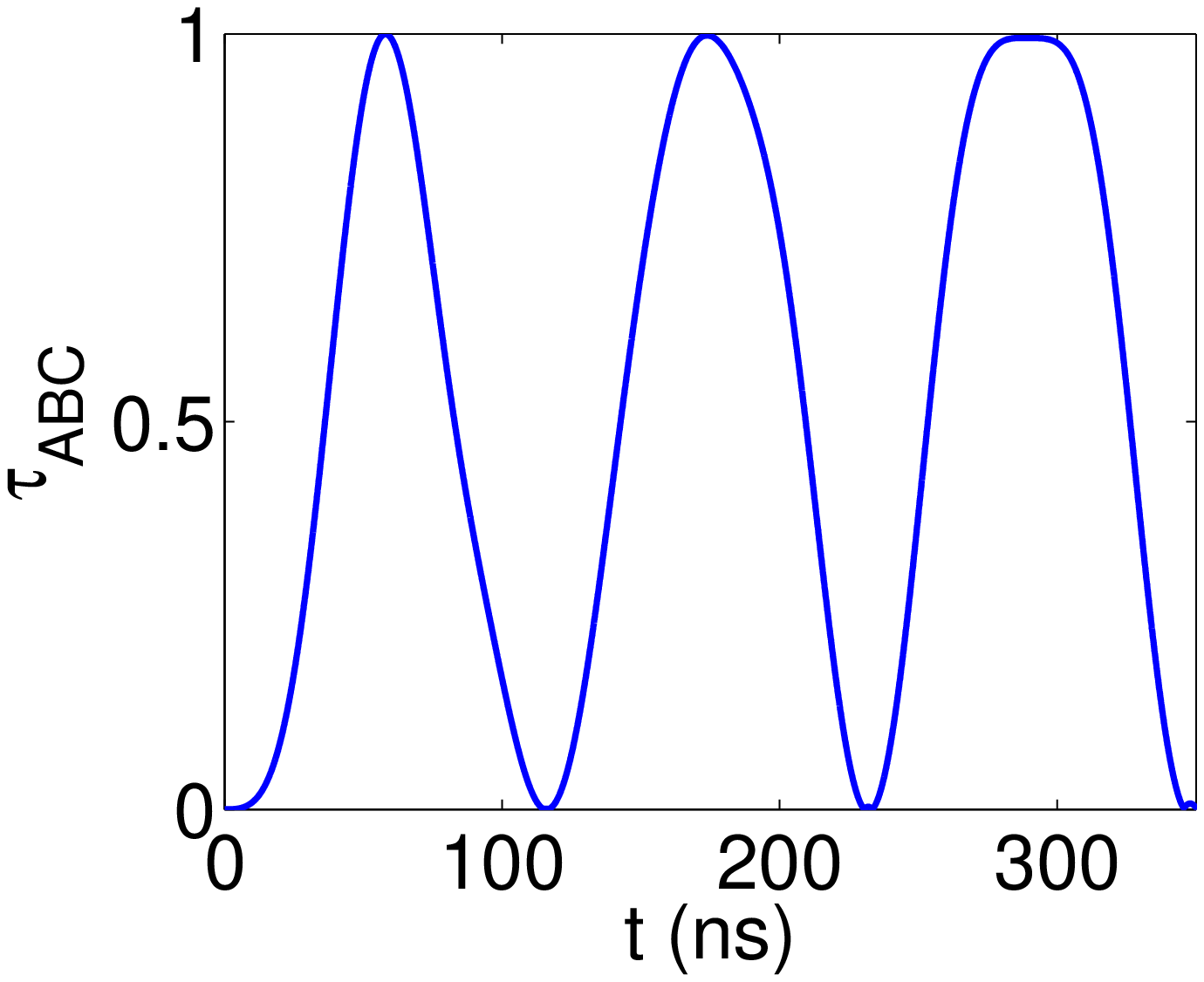}}
\subfigure{\includegraphics[width=.48\columnwidth]{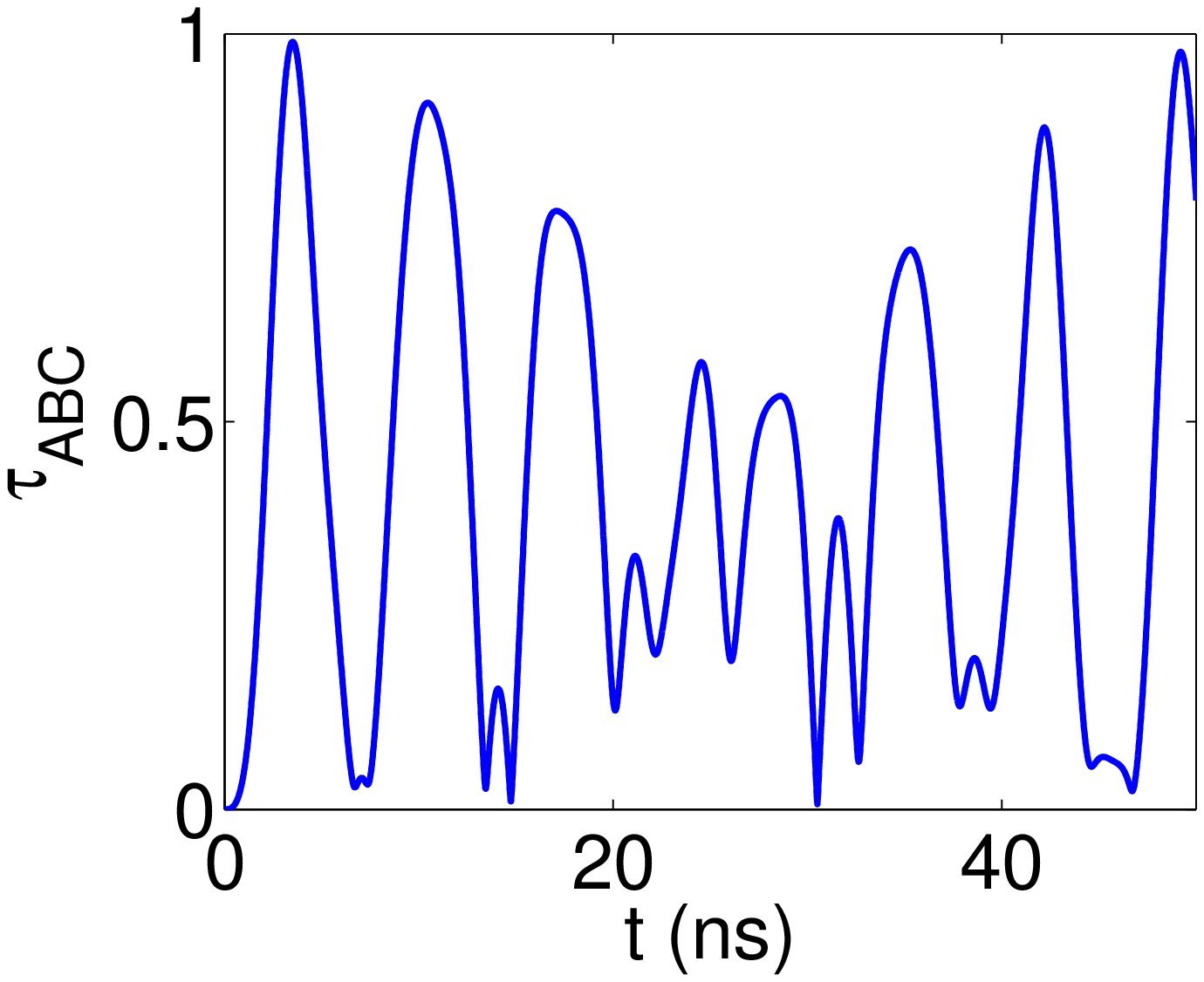}}
\subfigure{\includegraphics[width=.48\columnwidth]{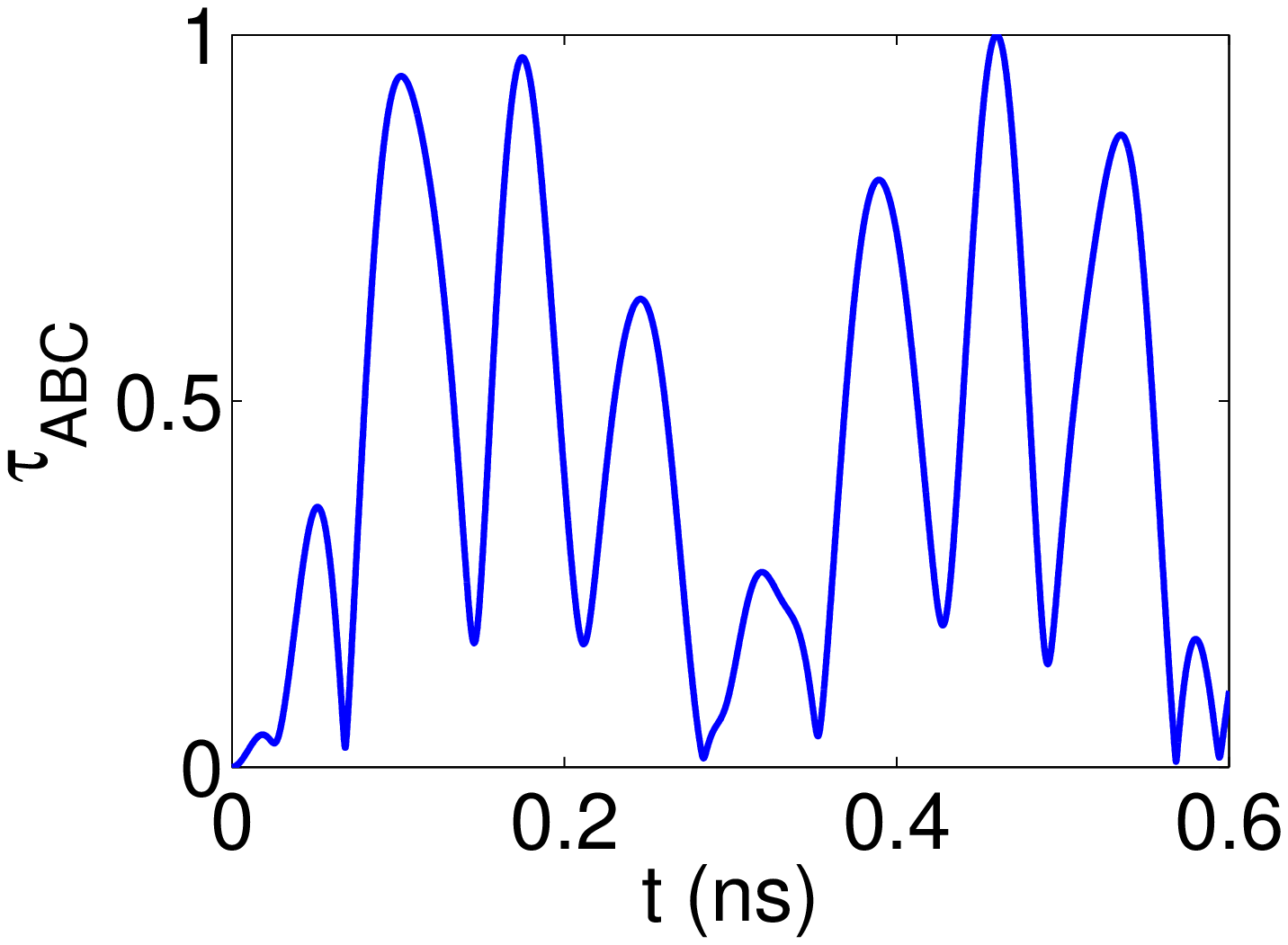}}
\subfigure{\includegraphics[width=.48\columnwidth]{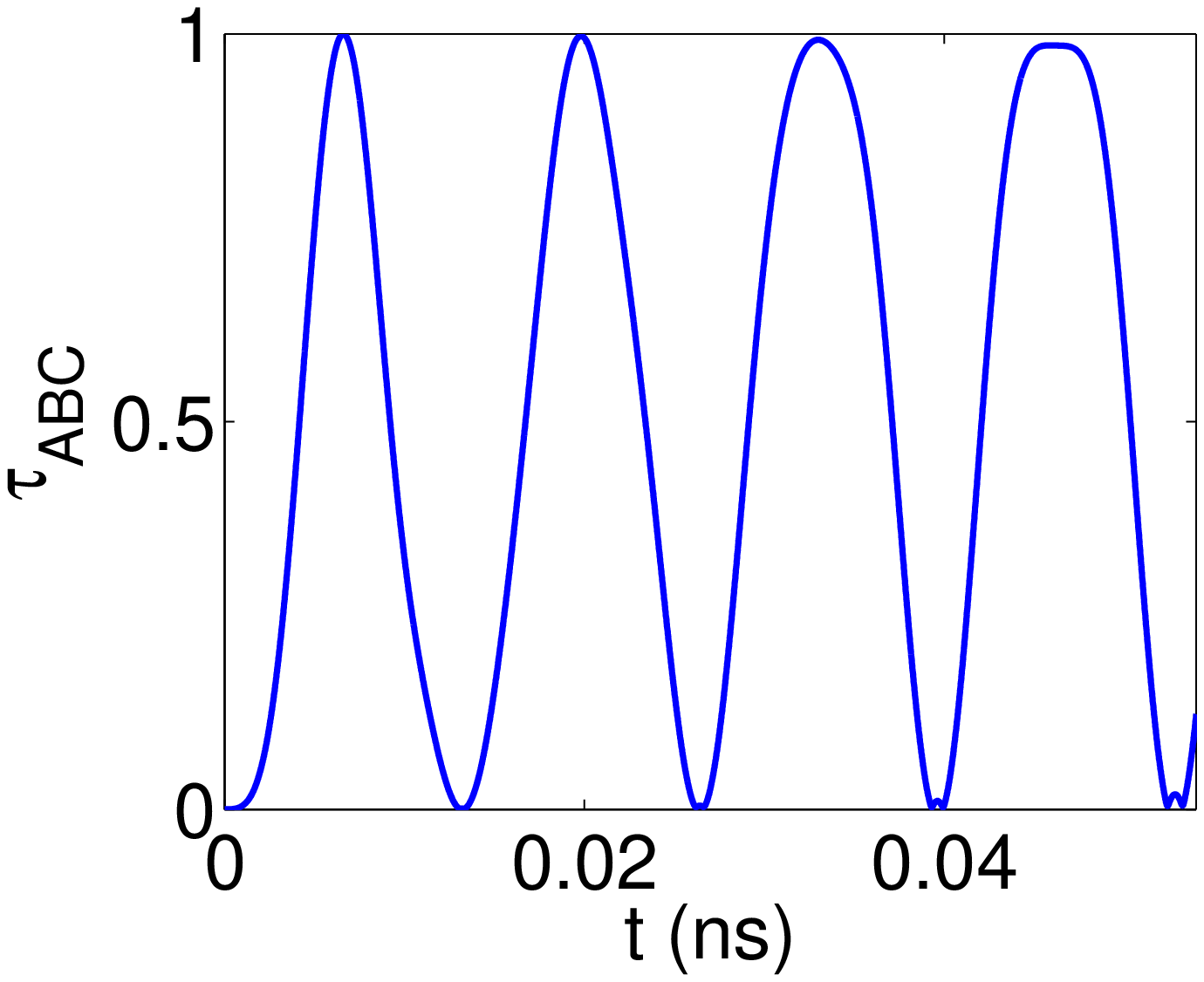}}
\caption{The evolution of the three-tangle with different detunings $\epsilon_A=\epsilon_B=\epsilon_C=\epsilon$ in the parallerly aligned case (the geometry is illustrated above the plots). At $t=0$, all qubits are initiated
in the $xy$-plane of the Bloch sphere. The qubits are then let to evolve, and the concurrences and the three-tangle are computed at each time step.
Upper left plot corresponds to $\epsilon=4.5$ meV, upper right to $\epsilon=4.7$ meV, lower left to $\epsilon=4.9$ meV, and lower right to $\epsilon=5.6$ meV.
}
\label{fig:cluster2}
\end{figure}

In the parallel alignment geometry (the illustration of the geometry is seen in Fig. \ref{fig:cluster2}), the sequential coupling scheme works similarly as in the linear case discussed in the previous section. The frequencies of the
entanglement oscillations are again determined by the pairwise $E_{cc}$-values. As the pairwise couplings are now of
the rectangular type, stronger couplings are achieved compared to the linear case, as can be seen in Fig. \ref{fig:delta_ddet}. The qualitative features of the entanglement oscillations in the sequential
scheme remain unchanged from the ones in Fig. \ref{fig:cluster}.

Although the Coulomb repulsion between the three qubits is still not symmetrical in the parallel formation, the differences between qubits
are smaller than in the linear case. This allows the generation of GHZ-states using a single detuning pulse applied simultaneously to all three qubits,
as in Ref. \onlinecite{maxtri}.
In this case, all qubits are again initiated in the $xy$-plane with the detunings $\epsilon_A=\epsilon_B=\epsilon_C=\epsilon$. The system is the evolved
with these constant detunings causing the qubits to entangle with each other. 
Fig. \ref{fig:cluster2} shows the tangle evolution with different detunings $\epsilon$.  

The value of the detunings, $\epsilon$, has a large qualitative effect on the $\tau_{ABC}$-oscillations. When $\epsilon$ is below or above the anti-crossing region
of the singlet charge states (i.e. $\epsilon<4.6$ meV or $\epsilon>5.5$ meV), $\tau_{ABC}$ oscillates between $0$ and $1$ with an approximately constant frequency.
As seen in the upper left and lower right plots of Fig. \ref{fig:cluster2}, the wave form is still not completely periodic due to the aforementioned asymmetry between the qubits. When the detuning is in the anti-crossing region (upper right and lower left plots in Fig. \ref{fig:cluster2}), the oscillations
are more complex with modulation-like behavior. Apart from the small asymmetry effects, the qualitative features of the tangle oscillations are similar to symmetric triangular case of Ref. \onlinecite{maxtri}, in which the behavior of the oscillations with different
detunings is also discussed with more detail.

The one-step preparation scheme depends in a more complex manner on the geometry of the $S-T_0$ qubit system than the sequential one.  
It could in principle be used in the linear geometry of Fig. \ref{fig:cluster} as well, by again setting the detunings to $\epsilon_A=\epsilon_B=\epsilon_C=\epsilon$ and letting the system evolve. However,
in this case, it is found to result in very irregular entanglement oscillations that would be ill-suited for the experimental creation of cluster states. The reason for this difference between the two
geometries is that in the linear geometry, the effect of the repulsion by the two other qubits depend heavily on the qubit in question. For example, the repulsion by A and B pushes the charge in C towards dot $6$, in this case effectively
lowering the $E_{cc}$-value between qubits B and C. On the other hand, the repulsion by B and C enhance the $E_{cc}$ value of A and B by pushing the charge in A into the dot $1$. In the parallel geometry, the repulsion between
the qubits is at its minimum, when all three qubits are in the $(1,1)$-charge state. The inter-qubit repulsion thus acts by inhibiting the transition to $(0,2)$. Although
the system is still not symmetric between the three qubits, the aforementioned effect is larger in B than in A and C, the effects of the asymmetry are much smaller than in the linear case. 

\subsection{General case and qubit arrays}

Finally, we will briefly discuss the creation of general cluster states between $S-T_0$ qubits. The generation of three-qubit GHZ-states was also studied with other intra-qubit orientations and distances, and the sequential entangling scheme was found
to work similarly to the linear case in all studied geometries. In some sense, the parallel case discussed in the previous section can be considered the optimal
geometry as it has the largest qubit-qubit couplings. The Coulomb-repulsion asymmetry between the qubits is also quite small in this formation. 

\begin{figure}[!ht]
\vspace{0.3cm}
\includegraphics[width=0.6\columnwidth]{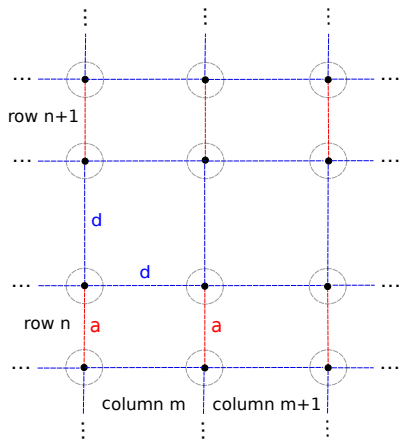}
\caption{(Color online) A two-dimensional $S-T_0$ qubit array. The intra-qubit dot distances are $a$ and the qubit qubit distances $d$. The 
array is divided into $N$ rows an $M$ arrays. In each row, the
qubits are coupled parallerly and in each column linearly.}
\label{fig:array}
\end{figure}

Extending the sequential entanglement scheme beyond three qubits (i.e. adding more qubits to the formations in Figs. \ref{fig:cluster} and \ref{fig:cluster2}) results
in an 1D-array of qubits with a corresponding cluster state. Coupling several of such 1D-arrays with each other then allows the creation general graph states.
Fig. \ref{fig:array} shows an illustration of a 2D-array $S-T_0$ qubits. In the figure, the 2D-array is divided in to $N$ rows and $M$ columns. In each row, the
qubits are coupled parallerly and in each column linearly. 

In the array geometry of Fig. \ref{fig:array}, it should in principle be possible to entangle all qubits in a single row into an cluster state with one
single detuning pulse (similarly as in Fig. \ref{fig:cluster2}). The neighboring rows can then be entangled with each other again sequentially in one step per a pair of rows.
Entangling the rows with each other should not affect the entanglement inside rows as the detuning values required for large $E_{cc}$-values in the linear and parallel alignments
differ considerably (see Fig. \ref{fig:delta_ddet}).
The total number of entangling steps for obtaining a cluster state spanning the whole array is thus $M+1$. A cluster state corresponding to a given graph in the array
can then be created by measuring the qubits that do not belong to the graph in the $\sigma_z$-basis, which effectively removes them from the cluster \cite{one-way}.

In reality, however, fabricating perfectly symmetrical qubit arrays is impossible. The asymmetries in the array probably inhibit the one-step preparation of cluster states for entire
rows of qubits. Instead, the preparation has to be done one qubit-pair at a time, or at least between row and column segments instead of
whole rows and columns. There are also many other experimental limitations, including how the electrostatic gating
defining the quantum dot potentials should be conducted in such arrays. The experimental issues are however outside the scope of this paper and the analysis in this section should
be considered a preliminary study on many-qubit arrays and cluster states in the $S-T_0$ qubit architecture. 

\section{Discussion}
 
Using exact diagonalization techniques, we have first computed the
energy eigenstates of a two-qubit system with several different qubit-qubit distances and discussed their effect on the capacitative coupling of the two-qubits. Longer
qubit-qubit distances were found to result in weaker coupling due to smaller differences in the Coulomb-repulsion between the qubit states.
The effect of the orientation of the qubits with respect to each other was also
discussed. We find that the coupling is strong with short qubit-qubit distances, and trapezoidal dot formations. These geometries are preferable in
creating efficient two-qubit gates. They allow smaller localization of the singlet electrons in the gate operation, which in turn decreases the charge-based decoherence and
charge state leakage between $S(1,1)$ and $S(0,2)$.

We also discussed the creation of cluster states and multi-qubit entanglement using the capacitative coupling.
Several inter-qubit geometries were studied. We simulated the creation of a
three-qubit cluster states using the extended Hubbard model with its parameters fitted to continuum model data. We simulated both the simultaneous and pairwise detuning schemes for the creation of three-qubit entanglement. 
We also discussed the creation of cluster states corresponding to large qubit arrays and arbitrary graphs.

In conclusion, we have studied the capacitative coupling of singlet-triplet qubits using exact diagonalization techniques. Our analysis on the geometry of the two-qubit
system and its effect on the coupling strength can be used to aid experimentalists in creating efficient realizations of the capacitative coupling scheme.
The analysis was also extended to three qubits, and the scheme for the creation of highly entangled cluster states should in principle be applicable to
any number of singlet-triplet qubits.

\section*{Acknowledgements}

We acknowledge the support from Academy
of Finland through its Centers of Excellence Program (project no. 251748).

\bibliography{lagrange}

\end{document}